\documentclass[prb,twocolumn]{revtex4}
\usepackage{graphicx}
\usepackage{hyperref}
\usepackage{amssymb}
\usepackage{dcolumn}
\usepackage{float}
\usepackage{booktabs}
\usepackage{amsmath} 
\usepackage[utf8]{inputenc}
\usepackage{lmodern}
\usepackage{color}
\definecolor{red}{rgb}{1.0,0.0,0.0}
\usepackage{makecell}
\usepackage{marginnote}

\newcommand{\nn}{\nonumber}

\newcommand{\e}{\mathrm{e}}

\DeclareMathAlphabet{\bi}{OML}{cmm}{b}{it}

\def\ba{\begin{aligned}}
\def\ea{\end{aligned}}
\def\be{\begin{equation}}
\def\ee{\end{equation}}
\def\bearr{\begin{eqnarray}}
\def\eearr{\end{eqnarray}}

\def\l{\left}
\def\r{\right}

\usepackage{tikz}

\begin{document}
\title{Unconventional phases in a Haldane model of dice lattice}
\bigskip
\author{Bashab Dey, Priyadarshini Kapri, Ojasvi Pal and Tarun Kanti Ghosh\\
\normalsize
Department of Physics, Indian Institute of Technology-Kanpur,
Kanpur-208 016, India}
\begin{abstract}

We propose a Haldane-like model of dice lattice analogous to graphene and explore its topological properties
within the tight-binding formalism. The topological phase boundary of the system is identical to that of Haldane
model of graphene but the phase diagram is richer than the latter due to existence of a distorted flat band. The
system supports phases which have a “gapped-out” valence (conduction) band and an indirect overlap between
the conduction (valence) band and the distorted flat band. The overlap of bands imparts metallic character to
the system. These phases may be further divided into topologically trivial and nontrivial ones depending on the
Chern number of the “gapped-out” band. The semimetallic phases exist as distinct points that are well separated
from each other in the phase diagram and exhibit spin-1 Dirac-Weyl dispersion at low energies. The Chern
numbers of the bands in the Chern-insulating phases are $0$ and $\pm2$. This qualifies the system to be candidate for
quantum anomalous Hall effect with two chiral channels per edge. Counterpropagating edge states emanate from
the flat band in certain topologically trivial phases. The system displays beating pattern in Shubnikov de Haas
oscillations for unequal magnitude of mass terms in the two valleys. We show that the chemical potential and
ratio of topological parameters of the system viz. Semenoff mass and next-neighbor hopping amplitude may be
experimentally determined from the number of oscillations between the beating nodes and the beat frequency,
respectively. 
\end{abstract}

\maketitle

\section{Introduction}
Engineering topological phases in materials has become an indispensable part of 
modern condensed matter physics. Although the notion of topology originated in mathematics long 
time back, it gained impetus from the discovery of Quantum Hall Effect (QHE)\cite{qhe}. 
QHE demonstrated that when a two-dimensional electron gas is subjected to strong magnetic field, 
the Hall resistance forms a series of plateaus quantized at $(h/\nu e^2)$ as 
the magnetic field or carrier density is varied. The number $\nu$ which defines 
the quantization may take integer (Integer QHE)\cite{qhe} or fractional values 
(Fractional QHE)\cite{fqhe}. 
The quantized effect was attributed to the formation of Landau levels 
\cite{laughlin,trugman,ilani,vasilo,tong} or magnetic Bloch bands 
\cite{tknn,thouless,simon,kohmoto,niu,girvin} in presence of a constant magnetic flux. 
Each of these bands may have a non-zero integer associated with it called the Thouless-Kohmoto-Nightingale-Nijs (TKNN) invariant.  
As long as the Fermi level lies in an energy gap, the Hall conductivity is given by the 
sum of invariants of all the bands lying below the Fermi level. The invariants resist 
any change from adiabatic perturbations in the system, which accounts for robustness 
of quantum Hall plateaus. The quantization has been predicted 
\cite{gusynin,peres,vasilo-graphene,vasilo-gapped,silicene,2dlandau,bi} and observed 
\cite{zhang,ziang,bilayer,phosphorus,inse,bise} in wide range of quasi-2D systems.

Although a constant flux appeared to be necessary to create Landau levels for the 
Hall quantization, it was proposed by F. D. M. Haldane that even a zero flux would 
do\cite{haldane}. 
In his model, Haldane considered a honeycomb lattice (graphene) with sublattice 
symmetry breaking potential and a periodic magnetic flux such that net flux linked 
with an unit cell vanishes. This breaks time-reversal symmetry (TRS) and inversion 
symmetry (IS) of the system without altering the original periodicity of the lattice. 
The phase space of sublattice potential and the periodic flux reveals the existence 
of a gapped phase, where the bands have non-zero TKNN invariants. It gives a quantized 
Hall conductivity similar to QHE when the Fermi energy lies in the gap. This phenomenon 
gave birth to the idea of Quantum Anomalous Hall Effect (QAHE). Breaking TRS is a necessary 
condition for QAHE to occur. Driving a system with high frequency circularly polarized light 
may also exhibit QAHE owing to the breaking of TRS\cite{kitagawa}. Several systems displaying 
QAHE have been fabricated recently\cite{qahewell,qahemag,mclver}.
The QHE and QAHE represent two distinct phenomena but they are unified by the concept of 
topology. These systems belong to the symmetry class A of the topological 
classification \cite{classification}. Under this class, each band of a 2D insulator 
has a uniquely defined topological invariant $\mathbb{Z}$ called Chern number associated 
with it, which is always an integer and is not protected by TRS, particle-hole or 
chiral symmetry. The quantized Hall conductances of these systems are directly related 
to the Chern numbers of the bands and are hence called Chern insulators. The Chern numbers 
of all bands identically vanish for a trivial insulator. 

Motivated by the possibility of new Chern phases, we propose a Haldane-like model 
\cite{haldane-dice1,haldane-dice2} of dice lattice 
\cite{Sutherland,Vidal,Korshunov,Rizzi,Wolf,JDMalcolm,Vigh} with broken sublattice symmetry 
and complex next nearest neighbour (NNN) hopping rendered by a staggered magnetic flux. 
Unlike graphene, we make a particular choice of a hexagonal unit cell where the staggered 
flux vanishes. This is necessary for drawing analogies with Haldane model of graphene. 
We compute the tight-binding band structure as a function of topological parameters such 
as Semenoff mass, NNN hopping and periodic flux. We get a phase boundary identical to that 
of Haldane model of graphene which separates the trivial and non-trivial topological phases. 
The phase diagram reveals the presence of metalic phases in addition to semimetalic, 
insulating and Chern insulating ones. The metalic phases are a consequence of indirect 
overlap between distorted flat band and conduction/valence band. The semimetalic phases 
are characterized by spin-1 Dirac-Weyl dispersion at either of the Dirac points 
and are represented by four distinct points in the phase diagram. The topological quantization 
in the Chern-insulating phases is twice as that of graphene. The quantization manifests itself 
as a pair of chiral edge states at either edge of a nanoribbon. 

The marriage of QHE and QAHE results in interesting phenomena like integer QHE 
in graphene\cite{haldane-mag}.
However, the behaviour of magneto-conductivity in quantum anomalous Hall systems 
remains unexplored.
The magneto-conductivity of a 2D electron system is known to exhibit Shubnikov de Haas (SdH) oscillations 
at strong magnetic fields and low temperature. In this work, we show that the Haldane 
model displays beats in the oscillations when the magnitude of mass terms in the two 
Dirac valleys are unequal and Fermi energy is close to higher Landau levels of the conduction 
or valence band. The beats can be used to extract information about the system parameters 
like Semenoff mass, NNN hopping and Fermi energy. Similar beating patterns have been observed 
in systems with Rashba spin-orbit coupling \cite{Winkler}.
 
This paper is organized as follows. In sec. II, we discuss about band structure
of a dice lattice with NNN hopping. The Haldane model of dice lattice, its phase
diagram and the anomalous Hall conductivity are discussed in sec. III. 
In sec. IV, edge states of Haldane-dice nanoribbon are presented. The beating pattern in
SdH oscillations of the Haldane-dice model subjected to the quantizing magnetic field
is presented in sec. V. Finally, summary of our results are presented in sec. VI.

\section{Dice lattice}
\begin{figure}[htbp]
\includegraphics[trim={1.5cm 2cm 0.5cm 1cm},clip,width=8cm]{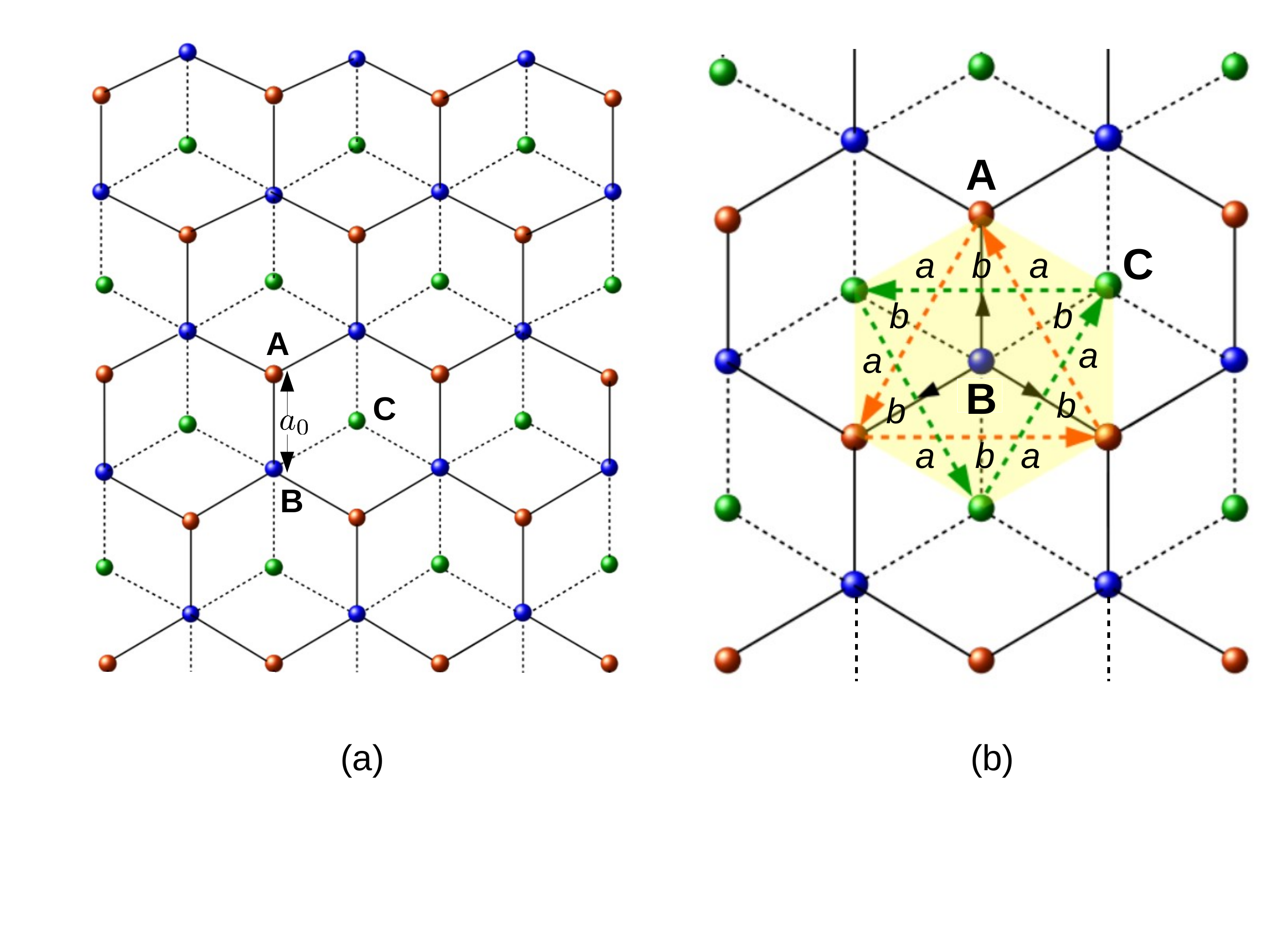}
\caption{(a) Sketch of a dice lattice. (b) A hexagonal unit cell (yellow-shaded) of 
the lattice with NN and NNN hoppings. The black arrows $({\bf a}_1,{\bf a}_2$ and ${\bf a}_3)$ represent the NN hopping vectors of B-type atom while the orange dotted $({\bf b}_1,{\bf b}_2$ and ${\bf b}_3)$ and green dotted $(-{\bf b}_1,-{\bf b}_2$ and $-{\bf b}_3)$
arrows are the NNN hopping vectors for 
A-type and C-type atoms . In Haldane model, a flux distribution is considered which has 
the same periodicity as that of the lattice subject to the condition that total flux 
through every unit cell vanishes. A symmetric flux distribution is considered such 
that the triangular regions $a$ (or $b$) formed by the paths of NNN hoppings have 
identical flux passing through them. }
\label{dicefig}
\end{figure}

The dice lattice is basically a honeycomb lattice with an additional atom at the
centre of each hexagonal unit cell from which the electron can hop only to atoms at
alternate vertices of the hexagon as shown in Fig. \ref{dicefig}(a). This leads to a
bipartitite lattice structure with two types of sites -- \textit{rim} sites (A and C)
and \textit{hub} sites (B) with coordination numbers  3 and 6 respectively.
The hopping amplitudes for nearest neighbour pairs A-B and B-C are identical
(say $t/\sqrt{2}$). The lattice has inversion symmetry with
\textit{hub} sites as the inversion centres. Dice lattice can be constructed
by growing trilayers of cubic lattices in [111] direction e.g. SrTiO${}_3$/SrIrO${}_3$/SrTiO${}_3$
heterostructure \cite{Ran}. An optical dice lattice may be generated by suitable
interference of three counter-propagating pairs of identical laser beams on a plane \cite{Rizzi}.


The dice lattice can also be thought of as a limiting case of 
$\alpha$-$\mathcal{T}_3$ lattice \cite{Rizzi,Raoux} with $\alpha=1$.
In recent years, there are several studies on diverse properties of
the dice lattice such as orbital susceptibility \cite{Raoux}, 
Klein tunneing \cite{Klein,Klein1}, zero-momentum optical
conductivity \cite{Illes,Illes1,Cserti,magneto-opt,non-linear,laser-opt}, 
magnetotransport properties \cite{Tutul,Malcolm,duan,Firoz}, magnetoplasmons 
\cite{mag-plasmon}, wave packet dynamics \cite{Tutul1}, 
electron states in the field of a charged impurity \cite{impurity,impurity1}, 
role of Berry phase in photoinduced gap, topological phase transition under Floquet
driving\cite{at3floquet,bdey}, effect of electromagnetic radiation on dice lattice 
\cite{radiation,radiation1},
electronic states of dice lattice ribbons 
\cite{dice-ribbon0,dice-ribbon,dice-ribbon1}, Ruderman-
Kittel-Kasuya-Yosida (RKKY) interaction \cite{RKKY} and chaotic dynamics
\cite{chaos}.

\begin{figure}[htbp]
\hspace{0cm}\includegraphics[trim={0cm 0cm 0cm 0cm},clip,width=9cm]{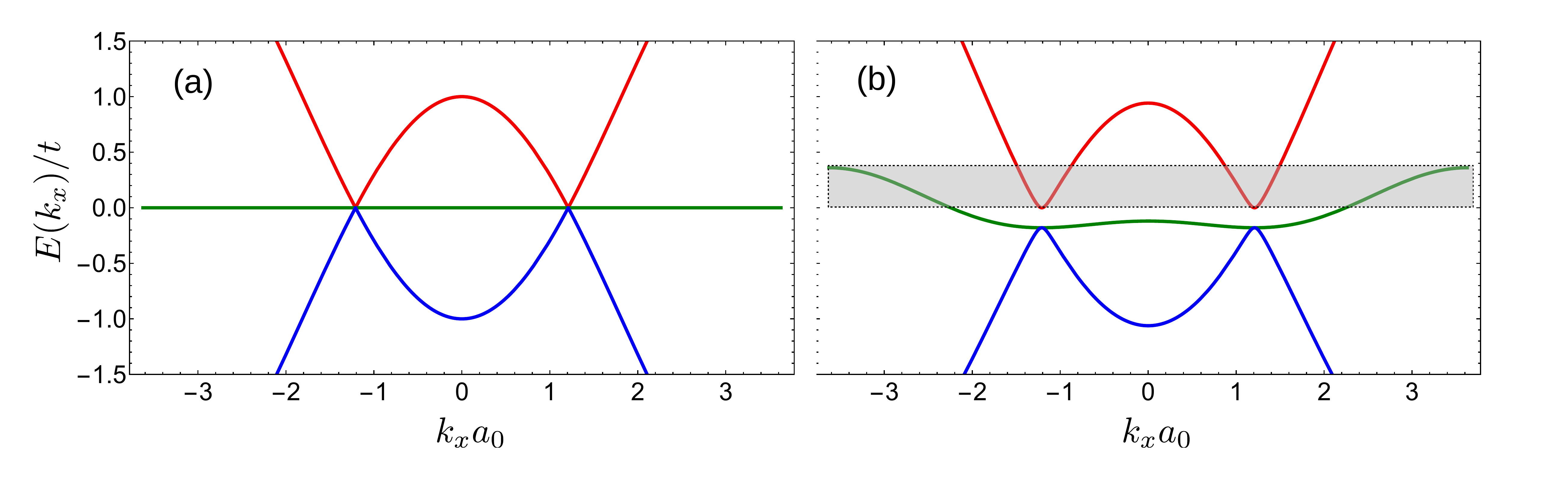}
\caption{Tight-binding bands of dice lattice with (a) $t_2=0$ and (b) $t_2/t=0.06$. 
The gray shaded region shows the indirect overlap between the distorted flat band and the 
conduction band. The bands are plotted along 
the line joining the high-symmetry ${\bf K^\prime, M }$ and ${\bf K}$ points.}
\label{dicebands}
\end{figure}

The tight-binding Hamiltonian of dice lattice in the basis of sublattices A, B and C is given as
\begin{equation}
H_0(\textbf{k}) = \frac{1}{\sqrt{2}}\left(\begin{array}
{ccc}
0 & t f(\textbf{k}) & 0 \\  
t f^*(\textbf{k}) & 0 & t f(\textbf{k}) \\ 
0 & t f^*(\textbf{k}) & 0
\end{array}\right),
\end{equation}
where ${\bf k} = (k_x, k_y)$,
$f(\textbf{k})=\sum_{j=1}^{3}\exp(-i{\bf k} \cdot {\bf a}_j)$, ${\bf a}_j$ are the nearest 
neighbour (NN) vectors as shown in Fig. \ref{dicefig}(b) and $t$ the NN hopping amplitudes. 
The explicit expressions of ${\bf a}_j$ are :--
${\bf a}_1=(\sqrt{3}/2,1/2)a_0$, ${\bf a}_2=(-\sqrt{3}/2,1/2)a_0$ and ${\bf a}_3=(0,1)a_0$ with
$a_0$ being the lattice constant.
The band structure comprises of a flat dispersionless band ($E_0=0$) flanked by two dispersive 
bands: $E_{\pm}=\pm t |f({\bf k})|$. The upper and lower bands are termed as conduction 
and valence bands respectively. The three bands touch each other with spin-1 Dirac-Weyl 
dispersion at two distinct points of the Brillouin zone ${\bf K}$ and ${\bf K}^\prime$ 
called Dirac points as shown in Fig. \ref{dicebands}(a). The low energy excitations 
around these points are governed by a pseudospin-1 Dirac-Weyl Hamiltonian given by
\begin{eqnarray}\label{ham-dice-low}
H_\mu({\bf q})=\hbar v_f(\mu q_x S_x+q_y S_y).
\end{eqnarray}
Here, $S_x$ and $S_y$ are the usual spin-1 matrices, $v_f$ the Fermi velocity and 
${\bf q}=(q_x,q_y)={\bf k-K}$ or ${\bf k-K^\prime}$. The index $\mu=+1$ and $-1$ represents 
${\bf K}$ and ${\bf K}^\prime$ valleys respectively. Diagonalising the Hamiltonian 
(\ref{ham-dice-low}), we get two linearly dispersive bands 
$E_\pm({\bf q})\equiv E_\pm(q)=\pm \hbar v_f q$ and the flat band $E_0=0$. 

Exact flat band, massless low energy excitations and three-fold degeneracy at Dirac points 
are rather approximate for this lattice. The band structure does not retain these features 
when NNN hoppings are taken into account. The NNN hopping amplitudes for A-A and C-C sites 
are identical by symmetry (say $t_2$). The B-B NNN hopping vanishes since it encounters 
the high potential barrier between A and C atoms. When the NNN hoppings are included, 
the Hamiltonian takes the form  
\begin{equation}
H(\textbf{k}) = \frac{1}{\sqrt{2}}\left(\begin{array}
{ccc}
2\sqrt{2}  \;  t_2  \;  d({\bf k})  & t f(\textbf{k}) & 0 \\  
t f^*(\textbf{k}) & 0 & t f(\textbf{k}) \\ 
0 & t f^*(\textbf{k}) & 2\sqrt{2} \;  t_2  \;  d({\bf k})
\end{array}\right),
\end{equation}
where $d({\bf k})=\sum_{j=1}^{3}\exp(i{\bf k} \cdot {\bf b}_j)$ where ${\bf b}_j$ are 
the NNN vectors as shown in Fig. \ref{dicefig}(b) with ${\bf b}_1=(\sqrt{3},0)a_0$, ${\bf b}_2=(-\sqrt{3}/2,3/2)a_0$ and 
${\bf b}_3=(-\sqrt{3}/2,-3/2)a_0$. Now, the bands are 
$E = 2 t_2 \; d({\bf k}) $ and 
$E_\pm=t_2 \; d({\bf k})\pm \sqrt{ t_2^2 d^2({\bf k})+t^2|f({\bf k})|^2}$. 
Expanding $d({\bf k})$ around the Dirac point ${\bf K}$ gives
\begin{equation}
d({\bf q-K}) = - \frac{3}{2} + \frac{9 a_0^2 q^2}{8} + \mathcal{O}(q^3).
\end{equation}
At the Dirac point ${\bf K}$, the eigen values are $-3t_2,-3t_2$ and $0$. 
This implies that a gap is created at the Dirac points reducing the three-fold degeneracy 
to two-fold, as shown in Fig. \ref{dicebands}(b). The flat band also becomes dispersive. 
Although there is a band gap, the conduction band states near the Dirac points overlap 
with those of the distorted flat band near the ${\bf \Gamma}$ point of the Brillouin zone. 
This indirect overlap imparts metalic character to the system even if the flat band is 
completely filled. 
Moreover, the band touching between the distorted flat band and the valence band is 
quadratic in first order. 

Graphene with only NN hopping also hosts two gapless bands with massless quasiparticles 
at the Dirac points. The inclusion of next nearest neighbour (NNN) hopping adds a 
{\bf k}-dependent scalar matrix to the tight-binding Hamiltonian in sublattice basis -- 
\begin{equation}
H_{t_2}^g(\textbf{k}) = \left(\begin{array}
{ccc}
 2 t_2  \; d({\bf k})  & t f(\textbf{k}) \\  
t f^*(\textbf{k}) & 2 t_2 \; d({\bf k})
\end{array}\right).
\end{equation}
So, at any Dirac point, $E_+=E_-=2 t_2 \; d({\bf K})=-3 t_2$. Thus, NNN hoppings only 
shift the Dirac points in graphene instead of opening up a gap \cite{goerbig}. 

\section {Haldane-like model of dice lattice}
We consider a spatially periodic magnetic flux through the plane of the lattice such that 
total flux through the hexagonal unit cell centred around any hub site (B) vanishes, as shown 
in Fig. \ref{dicefig}(b). Under such an orientation, the flux enclosed by hexagons formed 
by the paths of NN hoppings A-B or B-C also vanish by symmetry. Hence, the vector potential 
${\bf A(r)}$ can be chosen to vanish along those paths so that NN hoppings do not acquire 
Aharonov-Bohm phases ($\sim \int_{}^{NN}{\bf A(r)}\cdot d{\bf r}$). The flux enclosed by 
the triangle formed by NNN hoppings is non-zero. Hence, they aquire phases 
$\pm\phi_h\sim \pm\int_{}^{NNN}{\bf A(r)}\cdot d{\bf r}$ such that 
$t_2\rightarrow t_2 \e^{\pm i \phi_h}$. The sign of the phase is `$+$' for clockwise 
and `$-$' for counter-clockwise hopping. The value of $\phi_h$ is proportional to the 
flux enclosed by the three cyclic NNN hoppings A-A or C-C. On adding onsite energies $M$ 
(Semenoff mass) and $-M$ to A and C type atoms respectively, the lattice becomes a 3-level 
version of Haldane model. The Hamiltonian reads
\begin{eqnarray} \label{hamfinal}
\begin{aligned}
H({\bf k}) & = 2 t_2 h_0({\bf k})\cos\phi_h S_0 + (M- 2 t_2 h_z({\bf k})\sin\phi_h) S_z \\ 
& + t(g_x({\bf k})S_x+g_y({\bf k})S_y),
\end{aligned}
\end{eqnarray}
where $g_x({\bf k})=\sum_{i=1}^{3}\cos({\bf k} \cdot {\bf a}_i), \; 
g_y({\bf k})=\sum_{i=1}^{3}\sin({\bf k} \cdot {\bf a}_i),  
\; h_0({\bf k})=\sum_{i=1}^{3}\cos({\bf k} \cdot {\bf b}_i)$ and 
$h_z({\bf k})=\sum_{i=1}^{3}\sin({\bf k} \cdot {\bf b}_i)$. 
Also, $S_x,S_y,S_z$ are the usual spin-1 matrices and $S_0 \equiv$ diagonal matrix (1,0,1). 
The energy bands of Hamiltonian (\ref{hamfinal}) are obtained in Appendix [\ref{app-bands}].

On choosing the hexagonal unit cell centred around a {\it rim} site (A or B) with the flux 
orientation identical to that in Fig. \ref{dicefig}(b), the triangles formed by the NNN 
hoppings of the same {\it rim} site do not enclose any flux by symmetry. So, NNN hoppings 
of the corresponding {\it rim} site do not aquire any phase and may not be regarded as the 
conventional Haldane model. 

\begin{figure}[htbp]
\hspace{0cm}\includegraphics[trim={0cm 0cm 0cm 0cm},clip,width=6cm]{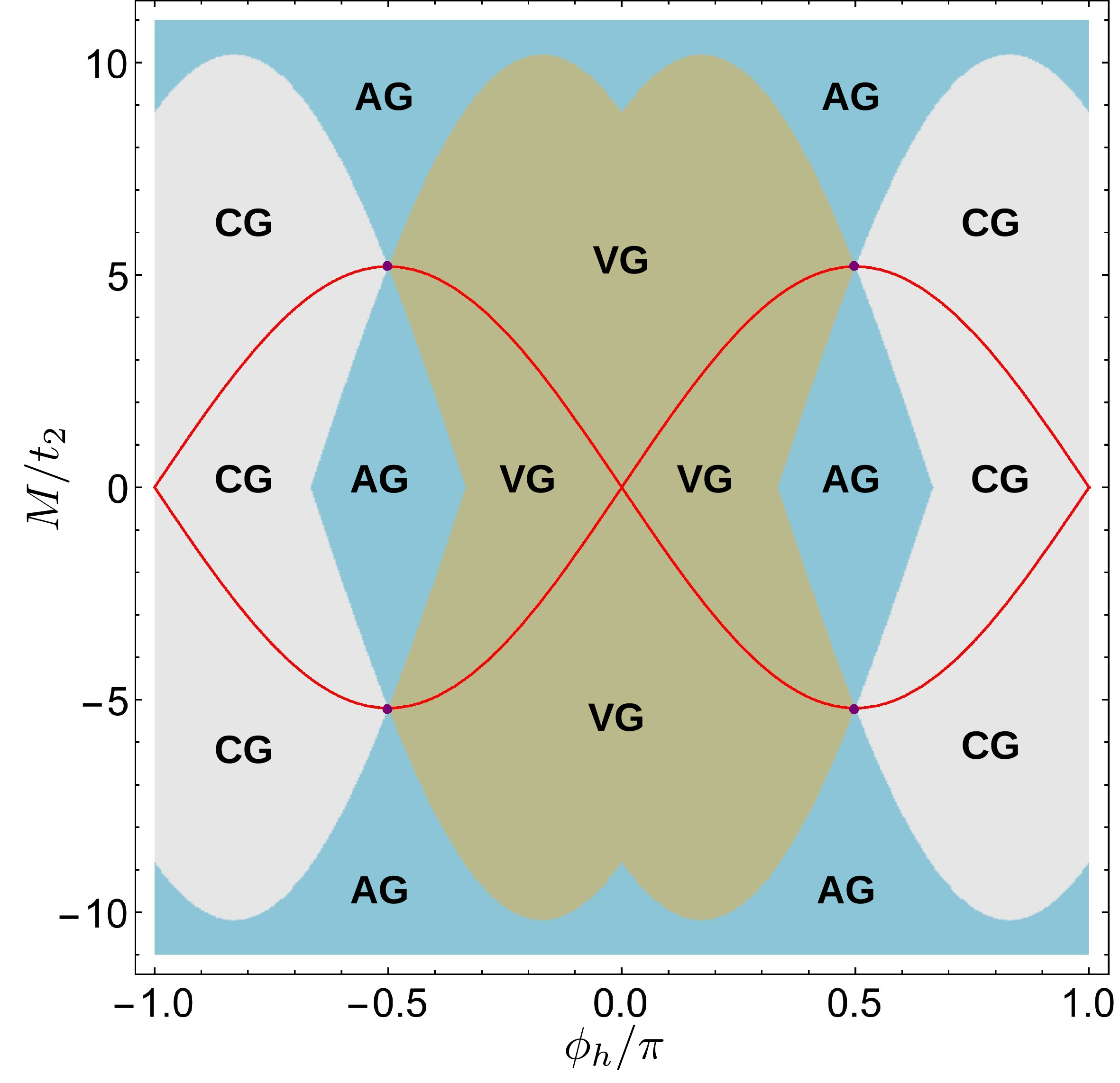}
\caption{Phase diagram of Dice-Haldane model.}
\label{dicephase}
\end{figure}

\begin{figure}[htbp]
\hspace{-0.5cm}\includegraphics[trim={0cm 0cm 0cm 0cm},clip,width=9cm]{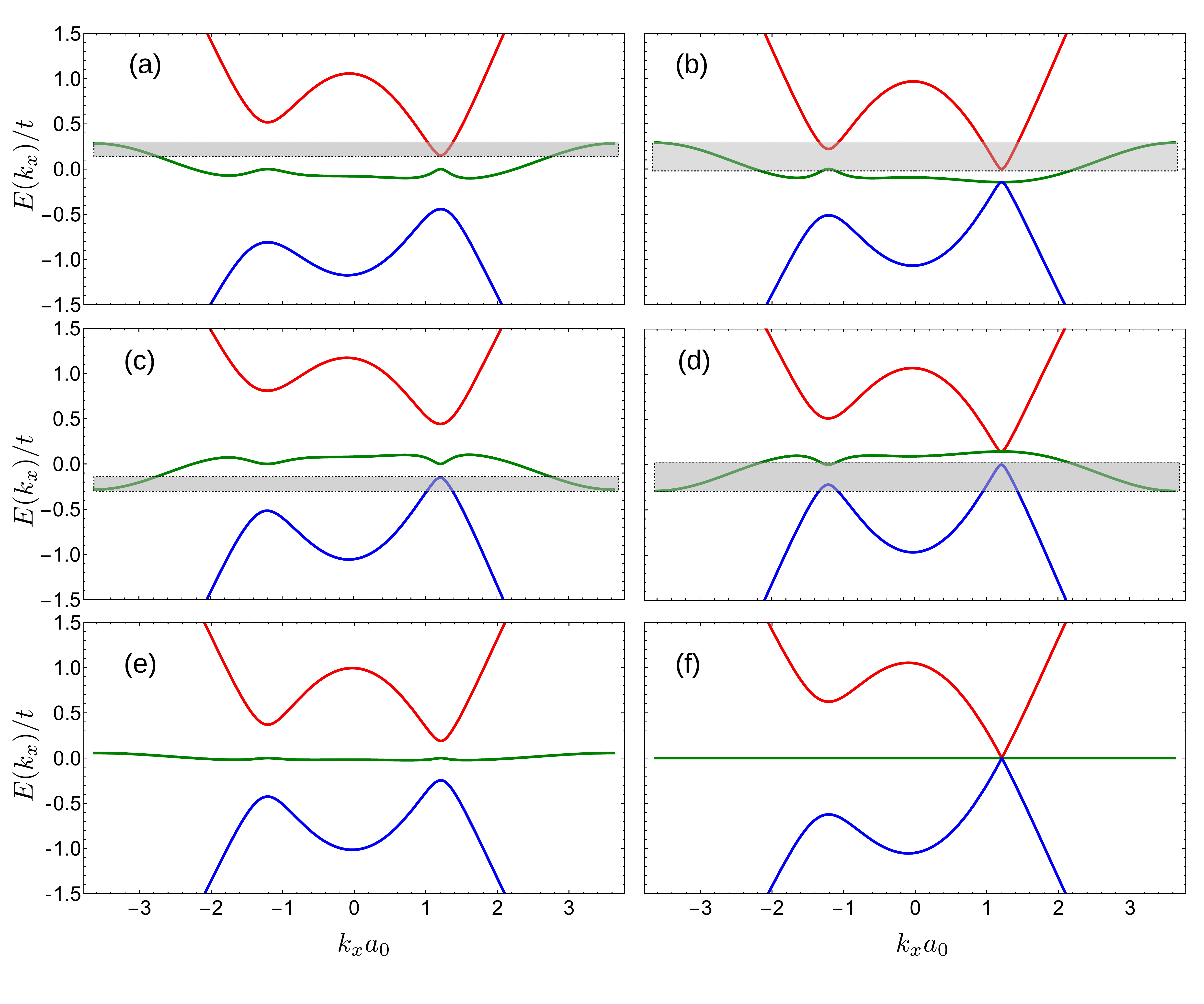}
\caption{Band structure of different phases in Haldane model of dice lattice :--
(a) A {\bf VG} phase, (b) a phase on the contour separating two {\bf VG} phases,
(c) a {\bf CG} phase, (d) a phase on the contour separating two {\bf CG} phases,
(e) an {\bf AG} phase and (f) a semimetalic phase.}
\label{phasebands}
\end{figure}

The phase diagram of the system governed by (\ref{hamfinal}) is shown in Fig. \ref{dicephase}. 
The region enclosed by the red curves represent topologically non-trivial phases while those 
outside it are trivial. The equations defining the red contours are $M=3\sqrt{3}\sin\phi_h$ 
and $M=-3\sqrt{3}\sin\phi_h$. The topological phase boundary is identical to that of Haldane 
model of graphene but there are several features which are in contrast. The topologically 
trivial and non-trivial phases are further divided into three categories -- {\bf VG}, {\bf CG} 
and {\bf AG} based on the band structure. The symbols {\bf VG}, {\bf CG} and {\bf AG} stand 
for `valence-gapped', `conduction-gapped' and `all-gapped' respectively. 
The `valence-gapped' means that valence band is gapped while the distorted flat and 
conduction bands have indirect overlap with each other [Fig. \ref{phasebands}(a)]. The overlap 
is similar to that in bare dice lattice with NNN hopping. The `conduction-gapped' implies that 
conduction band is gapped while the other two have indirect overlap [Fig. \ref{phasebands}(c)].
The `all-gapped' indicates that all bands are well separated from each other having no 
overlap at all [Fig. \ref{phasebands}(e)]. The red contour separates two {\bf VG} ({\bf CG}) 
phases because of closing and reopening of the band gap between the distorted flat and valence 
(conduction) bands along the contour  [Fig. \ref{phasebands}(b),(d)]. There are four 
independent purple points in the phase diagram at ($\pm0.5,\pm3\sqrt{3}$) where the conduction, 
flat and valence bands touch each other at either Dirac point with spin-1 Dirac-Weyl dispersion 
[Fig. \ref{phasebands}(f)].
The Chern numbers of the conduction, (distorted) flat and valence bands in the 
topologically non-trivial {\bf AG} phases around $\phi_h=\pm\pi/2$ are $\mp2,0$ 
and $\pm2$ respectively. In the non-trivial {\bf VG} and {\bf CG} phases for 
$\phi_h\lessgtr0$, the Chern numbers of the gapped valence and conduction bands 
are $\mp2$ and $\pm2$ respectively. The Chern numbers have been calculated using 
the discretized Brillouin zone method proposed by Fukui {\it et. al} \cite{fukui}. 
The system behaves as a Chern insulator when Fermi energy lies in a band gap of any 
of the of the topologically non-trivial phases. The system is metalic when Fermi 
energy lies in the range of overlapping bands in the topologically trivial as well 
as non-trivial {\bf VG} and {\bf CG} phases. The purple points can be termed as 
semimetalic when Fermi energy is at the three-fold band touching.
\begin{figure}[htbp]
\hspace{0cm}\includegraphics[trim={0cm 0cm 0cm  0cm},clip,width=9cm]{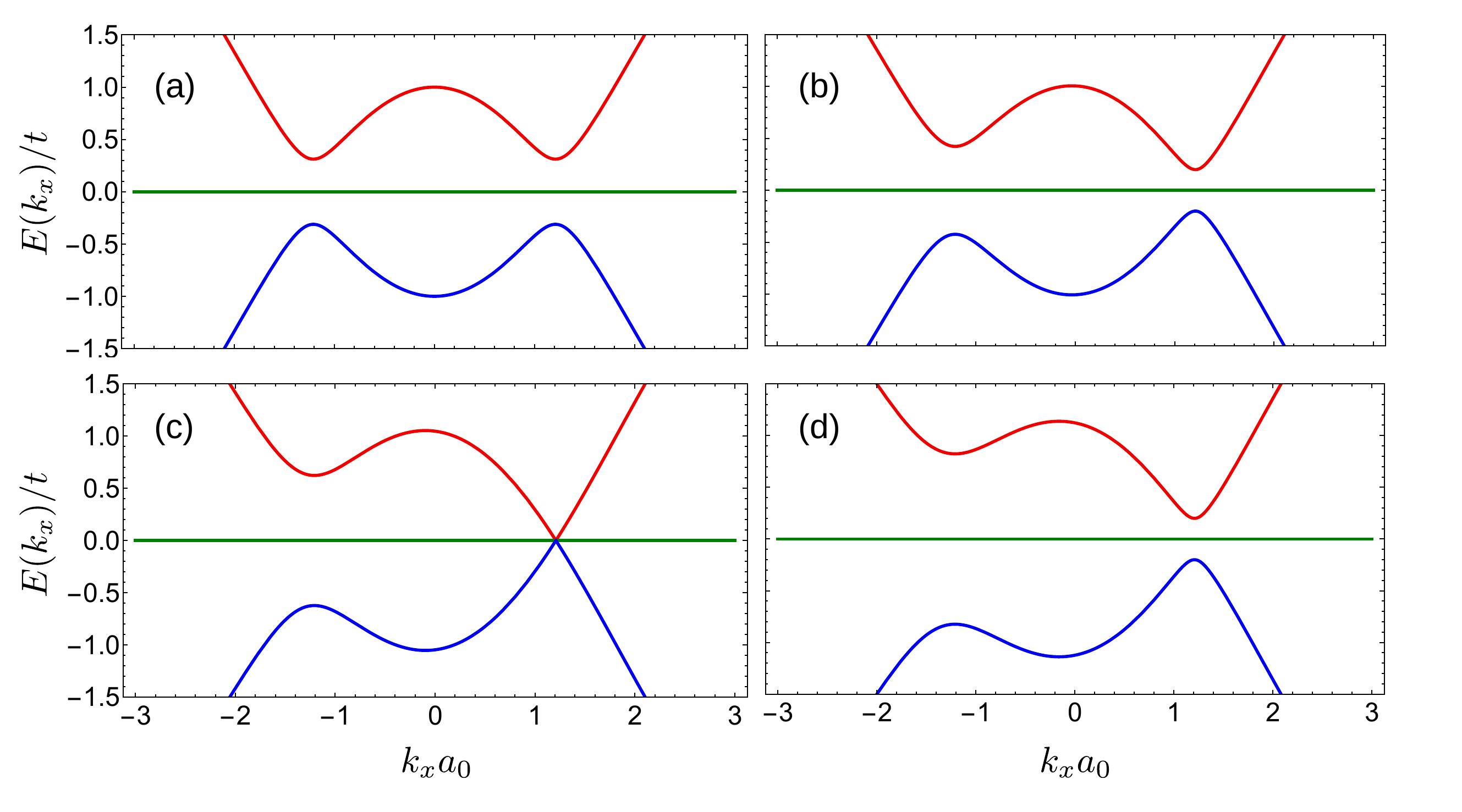}
\caption{Band structure of Haldane-dice lattice for $t_2/t=0.06, \phi_h=\pi/2$ and :--
(a) $M/t=0$, (b) $M/t=0.11$, (c)  $M/t=0.31$ and (d) $M/t=0.51$.}
\label{dicet2}
\end{figure}

For a particular choice of flux such that $\phi_h=\pi/2$, the Hamiltonian takes the form
\begin{eqnarray}\label{ham-piby2}
\begin{aligned}
H({\bf k})=(M-2 t_2 h_z({\bf k})) S_z+ t(g_x({\bf k})S_x+g_y({\bf k})S_y).
\end{aligned}
\end{eqnarray}
On diagonalizing, we get the bands $E_0=0$ and 
$ E_\pm = \sqrt{t^2 (g_x^2({\bf k})+g_y^2({\bf k}))+(M- 2 t_2 h_z({\bf k}))^2}$. 
Here $E_\pm$ are two dispersive bands symmetrically gapped around a zero energy flat 
band $E_0$ as shown in Figs. \ref{dicet2}(a)-(d)]. In this case, the flat band remains 
completely unperturbed by $t_2$. Thus, the flat band which becomes dispersive on inclusion 
of $t_2$ in bare dice lattice regains its flatness under the application of Haldane flux 
with $\phi_h=\pi/2$. In fact, a completely flat band occurs for $\phi_h=(2n+1)\pi/2$ where 
$n$ is an integer. Considering the symmetry in the band structure coming from pure imaginary 
NNN hoppings, we will consider the Haldane-dice model only with $\phi_h=\pi/2$ (and $M>0$) 
throughout the rest of the paper.

On linearizing the Hamiltonian (\ref{ham-piby2}) around the Dirac points 
${\bf K}$ and ${\bf K}^\prime$, we get
\begin{eqnarray}\label{ham-low}
H_\mu({\bf k})=\hbar v_f(\mu q_x S_x+q_y S_y)+m_\mu v_f^2S_z ,
\end{eqnarray}
where $\mu$ represents valley index, ${\bf q}$ is a small momentum vector 
w.r.t a Dirac point, $v_f=3 a_0 t/ 2\hbar$ and $m_\mu v_f^2=(M-\mu\epsilon_{t_2})$ with 
$\epsilon_{t_2}=3\sqrt{3} t_2$. The Hamiltonian (\ref{ham-low}) is analogous to that of massive 
spin-1 Dirac quasiparticles in two dimensions. The low energy bands are $E_0({\bf q}) = 0$ and
\begin{equation}\label{low}
E^\mu_\pm({\bf q}) = \pm \sqrt{(\hbar v_f q)^2+(m_\mu v_f^2)^2}.
\end{equation}
The valley-symmetry of the band structure is not preserved due to breaking of TRS and 
inversion symmetry. The band gaps at ${\bf K}$ are smaller than at ${\bf K}^\prime$.

The z-component of Berry curvature of the bands around these points are
\begin{equation}\label{berry}
\Omega_{\pm}^\mu({{\bf q}}) = \pm \mu 
\l[\frac{m_\mu \hbar^2 v_f^4}{(\hbar^2 v_f^2 q^2 + m_\mu^2 v_f^4)^{\frac{3}{2}}}\r], 
\hspace{0.5 cm}\Omega_{0}^\mu({\bf q}) = 0.
\end{equation}
\begin{figure}[htbp]
\includegraphics[trim={0cm 0cm 0cm 0cm},clip,width=6.5cm]{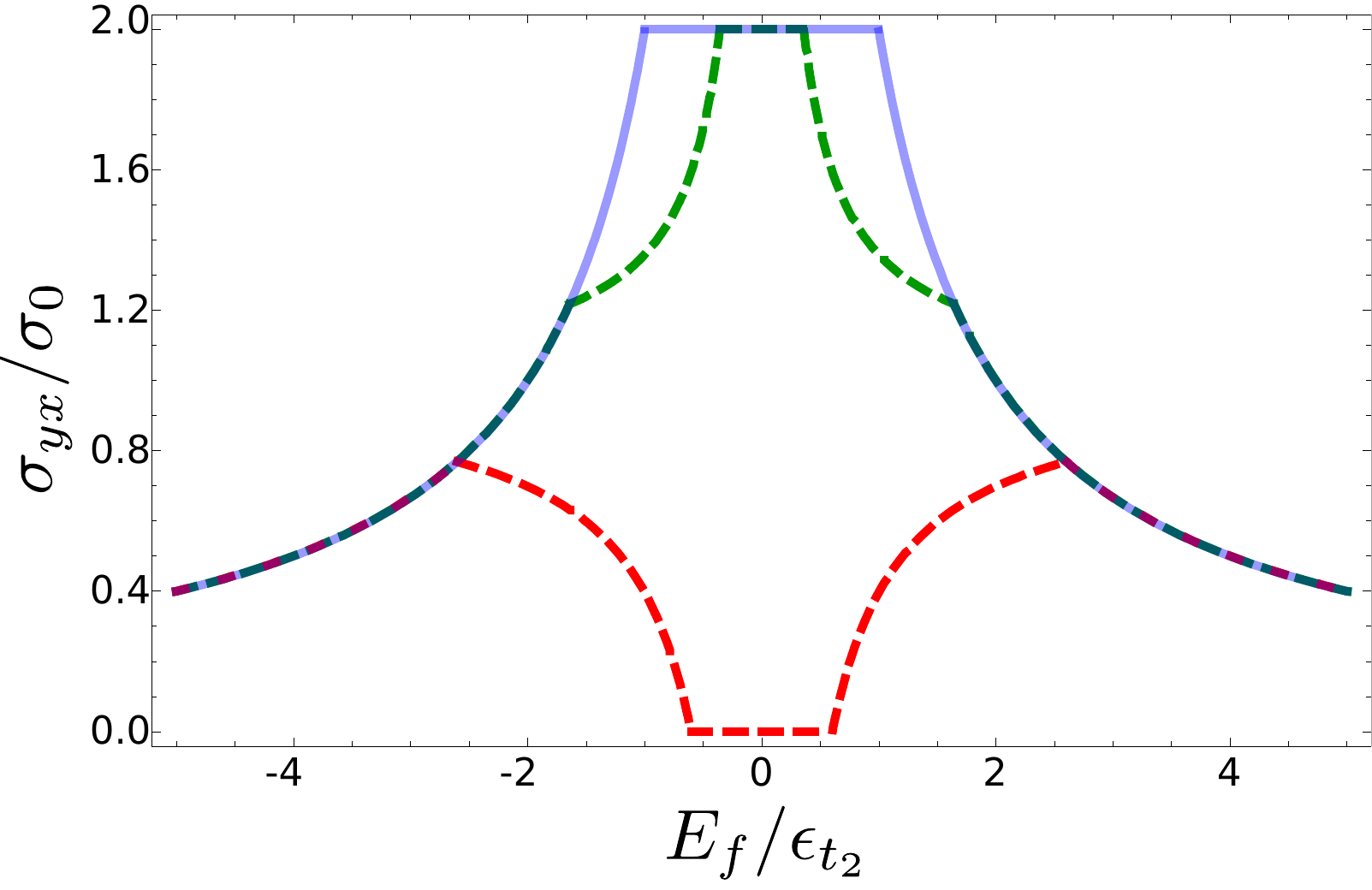}
\caption{Variation of the Hall conductivity ($\sigma_{yx}$) with Fermi energy ($E_f$) for
$\gamma=0$ (solid blue curve), $\gamma=1.6$ (lower, red dashed
curve) and  $\gamma=0.64$ (upper, green dashed curve).}
\label{hall}
\end{figure}
The anomalous Hall conductivity is given by\cite{Xiao-niu}
\begin{equation} \label{hallberry}
\sigma_{yx}(E_f) = \sigma_0 \sum_{\lambda, \mu}^{}\int_{0}^{\infty} 
\Omega_{\lambda}^\mu(q)f^\mu_\lambda (E_f)qdq,
\end{equation}
where $\lambda = 0,\pm 1$, $\sigma_0 = e^2/h$ and 
$f^\mu_\lambda (E_f) = [\e^{(E_{\lambda}^\mu-E_f)/k_BT}+1]^{-1}$ 
is the Fermi-Dirac distribution function.
Using Eqs. (\ref{berry}) and (\ref{hallberry}), the Hall conductivity of the model at 
$T=0$ as a function of Fermi energy $E_f$ is obtained as -- \\
\underline{Case I: $\gamma<1$}
\begin{equation}
\begin{aligned}\label{ml1}
\frac{\sigma_{yx}(\epsilon_f)}{\sigma_0} = &  \l(\frac{2}{|\epsilon_f|}\r) 
\theta[|\epsilon_f|-(\gamma+1)] + \l(1+\frac{1-\gamma}{|\epsilon_f|}\r) \\ 
& \theta[|\epsilon_f|-|\gamma-1|]\theta[|\gamma+1|-|\epsilon_f|] + \\ 
& \l(2\r)\theta[|\gamma-1|-|\epsilon_f|],
\end{aligned}
\end{equation}
\underline{Case II: $\gamma>1$}
\begin{equation}
\begin{aligned}
\frac{\sigma_{yx}(\epsilon_f)}{\sigma_0} = &  \l(\frac{2}{|\epsilon_f|}\r) 
\theta[|\epsilon_f|-(\gamma+1)] + \l(1+\frac{1-\gamma}{|\epsilon_f|}\r) \\ 
& \theta[|\epsilon_f|-|\gamma-1|]\theta[|\gamma+1|-|\epsilon_f|],
\end{aligned}
\end{equation}
\underline{Case III: $\gamma=0$}
\begin{equation}
\begin{aligned}
\frac{\sigma_{yx}(\epsilon_f)}{\sigma_0} = &  \l(\frac{2}{|\epsilon_f|}\r) 
\theta[|\epsilon_f|-(\gamma + 1)] + 
\\ & \l(2\r)\theta[|\gamma-1|-|\epsilon_f|],
\end{aligned}
\end{equation}
where $\epsilon_f = E_f/\epsilon_{t_2}$ and $\gamma = M/\epsilon_{t_2}$. 
The variation of Hall conductivity of the system with $E_f$ is shown in Fig. \ref{hall} 
for different values of $\gamma$. For $\gamma=0$, $\sigma_{yx}$ varies smoothly as 
$\sim 1/|E_f|$ when $E_f$ is below or above the band gap due to a valley-symmetric band structure. 
For $0<\gamma<1$ and $\gamma>1$, cusps appear in $\sigma_{yx}$ when $E_f$ enters or leaves the 
gap at ${\bf K}^\prime$ point due to asymmetry of band structure in the two valleys. 
For $\gamma<1$, the Chern number of valence and conduction bands are 2 and -2 respectively 
while that of flat band is zero. For $\gamma>1$, the Chern numbers of all the bands vanish 
and it acts like a trivial insulator. Thus, the Hall conductivity is quantized as 
$2\sigma_0$ and vanishes to 0 for $\gamma<1$ and $\gamma>1$ respectively when $E_f$ lies 
in the bulk band gap (at ${\bf K}$). 


Dice lattice also hosts a Floquet topological phase identical to the case of Haldane 
model with $M=0$ and $\phi_h=\pi/2$, when shine with circularly polarized light \cite{bdey}. 
A similar result was obtained for monolayer graphene where the commutator in the 
effective Floquet Hamiltonian in real space is equivalent to the second nearest neighbour 
hopping with phase $\pi/2$ of Haldane model \cite{kitagawa}.

\section{Edge states of Haldane-Dice nanoribbon}

\begin{figure}[htbp]
\hspace{0cm}\includegraphics[trim={0cm 2cm 0cm 2cm},clip,width=9cm]{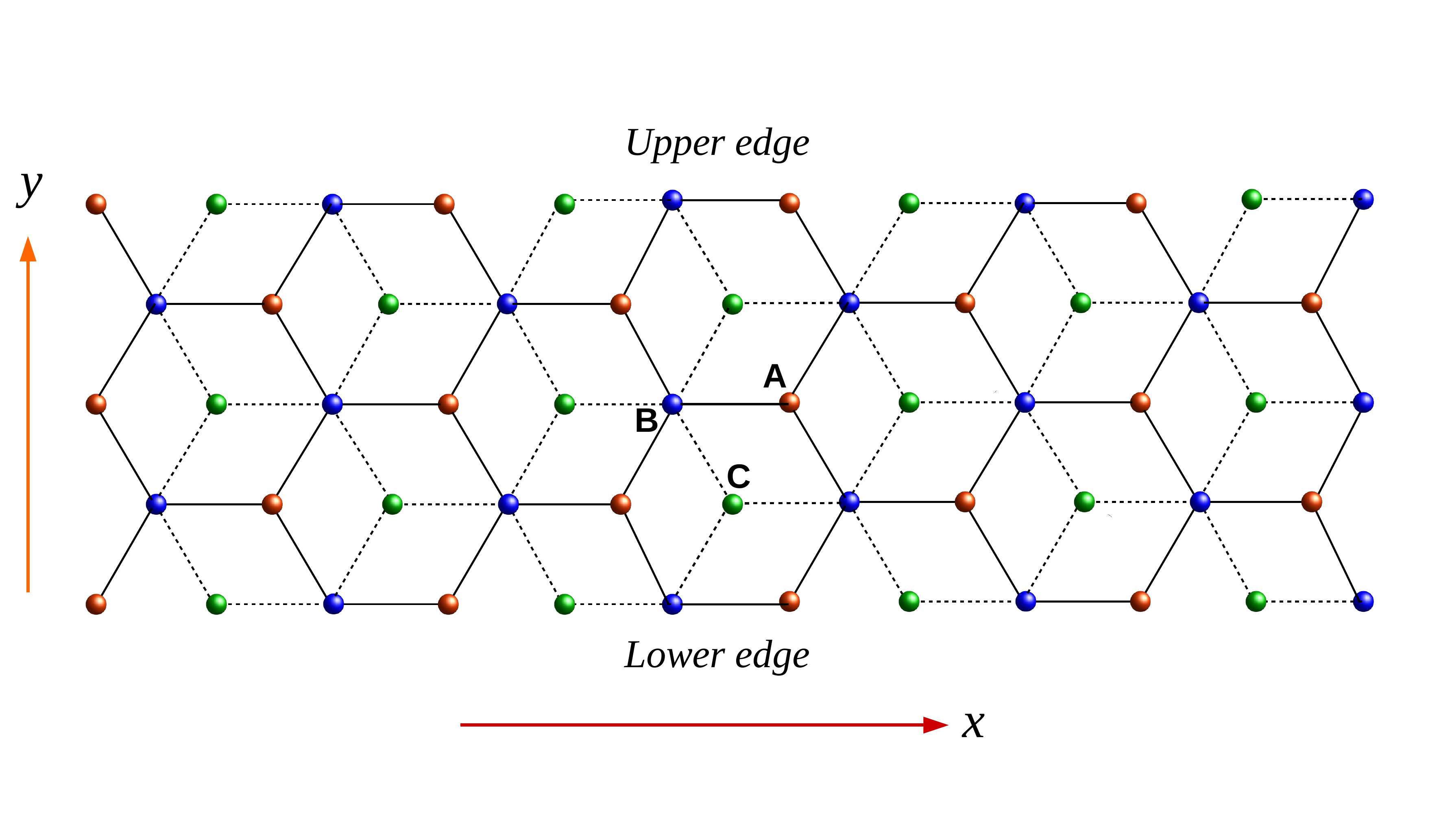}
\caption{Schematic diagram of a dice nanoribbon with armchair edges.}
\label{dicestrip}
\end{figure}

\begin{figure*}[htbp]
\hspace{-0.5cm}\includegraphics[trim={0cm 0cm 0cm  0cm},clip,width=18cm]{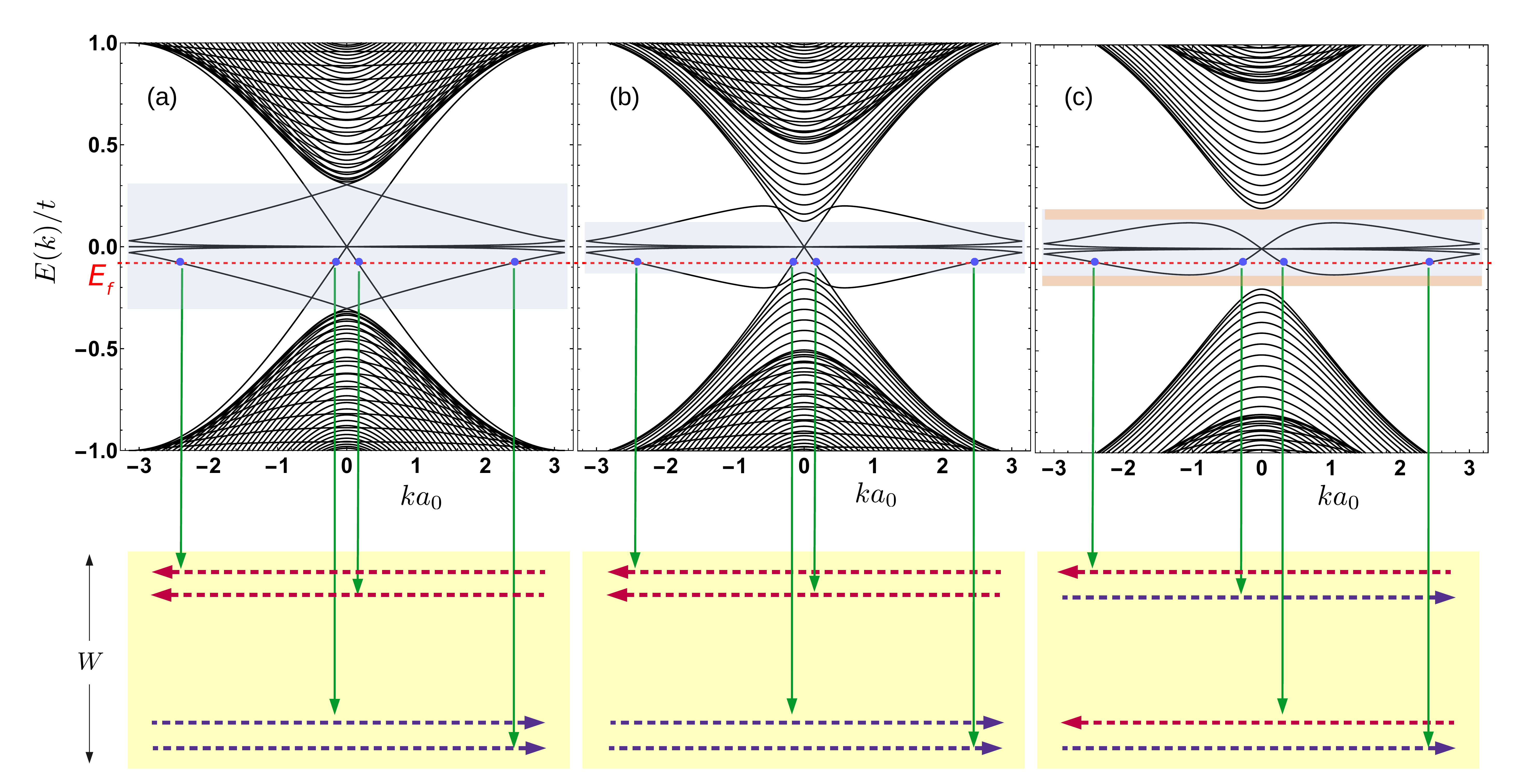}
\caption{Low energy tight-binding bands of Haldane-dice nanoribbon for $\phi_h=\pi/2,\; t_2/t=0.06$ and 
(a) $\gamma=0$, (b) $\gamma=0.64 $ and (c) $\gamma=1.6$, where $k=3k_x$. Schematic sketches of a 
part of the infinite nanoribbon are shown below the plots as yellow rectangles. 
The width $W$ of the nanoribbon = 41 hexagonal cells $\approx 71 \;a_0$ . 
There are four edge states (blue dots) at the Fermi energy $E_f$ (dashed red line). 
For $\gamma<1$ i.e. the topologically non-trivial regime, there exists two chiral states at 
either edge [Figs. (a),(b)].  For $\gamma>1$ i.e. the topologically trivial case, the edge 
states are counter-propagating at either edge [Fig. (c)].}
\label{diceedge}
\end{figure*}

The calculation of the Chern number requires evaluation of Berry curvature 
$\Omega_{k_x k_y}$ and its integration over the 2D Brillouin zone. Hence, 
a two-component parameter space is mandatory for the concept of Chern number.
However, mesoscopic measurements are done on narrow strips of a material. The finite 
width of the strip acts as a confining potential which breaks the periodicity of latttice  
along the confining direction and allows propagating states only in the direction perpendicular 
to it. The 2D bands of an insulator decompose into a set of densely spaced 1D sub-bands 
\cite{supriyo} representing bulk states, with gaps in the spectra. For a Chern insulator, 
there exists a set of states in the gaps which connect two adjacent bulk bands. 
The wave functions of these states decay exponentially from the edge of the strip towards 
its bulk. They propagate along a fixed direction at either edge and hence are called 
chiral edge states. Due to their chiral nature, the edge states do not undergo backscattering 
from impurities and carry a dissipationless current. When the Fermi energy lies in a band gap, 
only the dissipationless edge states in the gap conduct, thereby giving rise to 
quantized Hall plateaus and nearly vanishing longitudinal resistance. The number of 
chiral edge states at the Fermi energy equals the sum of Chern numbers of all the 
2D bulk bands below it. Thus, the bulk topological invariants manifest themselves 
as chiral edge states. This is called bulk-edge correspondence \cite{hatsugai}.

We consider an armchair nanoribbon of the Haldane-dice lattice infinitely long along $x$ 
direction but having a finite width along $y$ as shown in Fig. \ref{dicestrip}. 
The tight-binding band structure of this strip is plotted for different values of 
$\gamma$ in Fig. \ref{diceedge}. The blue dots represent the edge states at a given 
Fermi energy $E_f$ within the bulk band gap (shaded light blue). For $\gamma<1$, there 
are two chiral modes (unidirectional) confined at either edge at a given energy in the bulk 
gap, as shown in Figs. \ref{diceedge}(a),(b). The velocities of the states at opposite 
edges are directed opposite to each other, thereby making them chiral. These states 
are responsible for quantized Hall conductance of $2e^2/h$ (neglecting spin) when $E_f$ 
lies in the bulk gap. This is consistent with Eq. (\ref{ml1}) by bulk-edge correspondence. 
For $\gamma>1$, there exists no state within a particular range (orange shaded) of the 
bulk energy gaps, as shown in Fig. \ref{diceedge}(c). So, the system will be a trivial 
insulator if $E_f$ lies in that range. However, there are counter-propagating edge states 
at either edge at energies close to flat band. These states exist even for $\gamma\gg1$. 
This kind of edge states are absent in Haldane model of graphene in the non-topological regime 
which implies that they are peculiar to pseudospin-1 Dirac-Weyl system and arise because 
of the flat band. Due to counter propagation, the pair of edge states will not carry a 
net charge current at either edge and the system will behave as an insulator.
Hence, bulk-boundary corresponds holds good in this case as well.

\begin{figure}[htbp]
\hspace{0cm}\includegraphics[trim={1cm 0cm 0cm 0cm},clip,width=9cm]{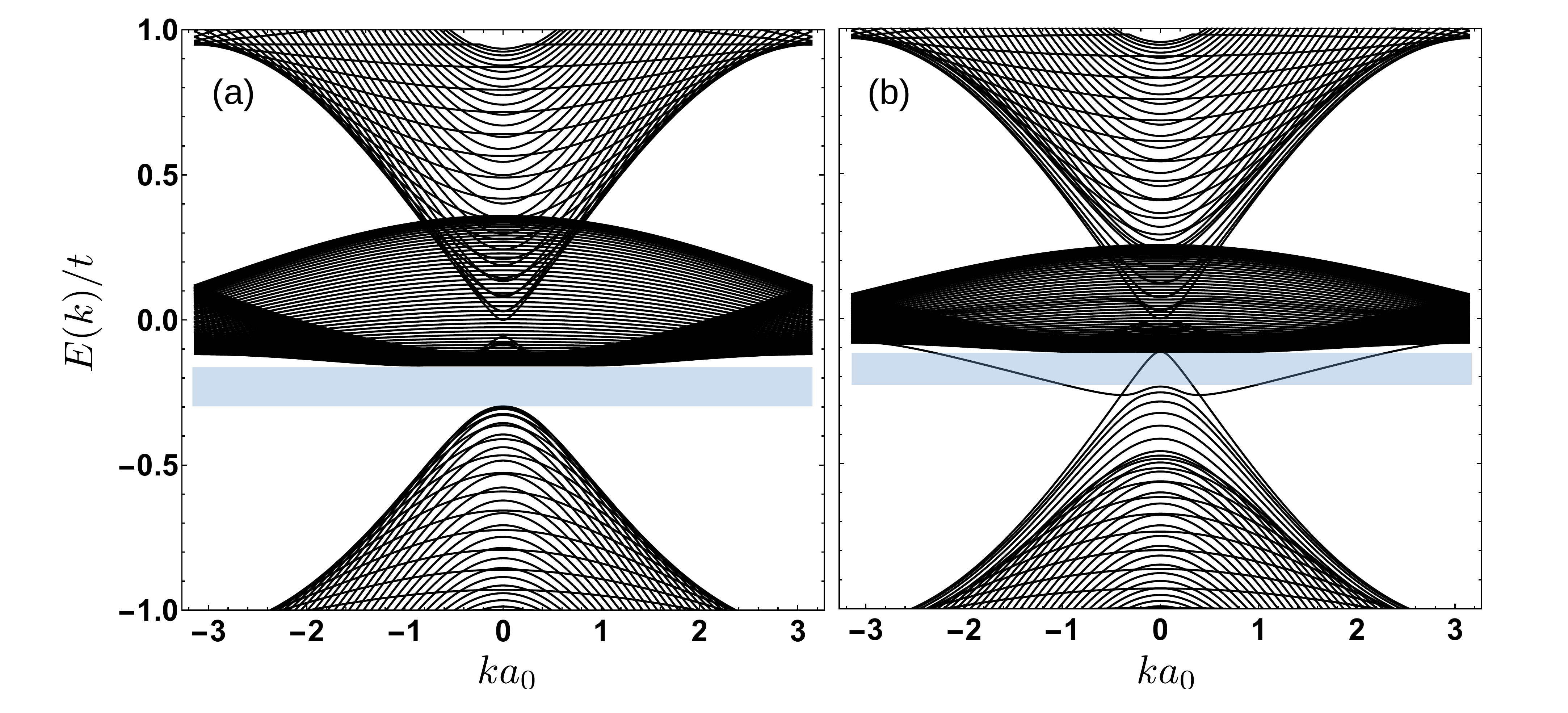}
\caption{Tight-binding bands of the Haldane-dice nanoribbon in (a)
topologically trivial [$\phi_h = 0, M=2t_2$] and (b) topologically non-trivial
[$\phi_h=0.25, M=2t_2$] {\bf VG} phases.}
\label{dicestripVG}
\end{figure}

The band structure of the nanoribbon in topologically trivial and
non-trivial {\bf VG} phases are shown in Fig. \ref{dicestripVG}(a) and 
Fig. \ref{dicestripVG}(b) respectively. In the trivial phase, no state 
exists in the bulk band gap
(shaded light blue) while chiral edge states fill the bulk gap in the
non-trivial regime. This testifies the topological nature of the system
despite the 'metalic' overlap between two bands. Similar edge states
appear for non-trivial {\bf CG} phases as well.

The armchair nanoribbon whose band structure is shown in Fig.(\ref{diceedge}) and (\ref{dicestripVG}) has number of atomic rows $N_r=83$ which is equal to $3N-1$ with $N=28$. It is known that the spectrum for an armchair nanoribbon is semimetalic for $N_r=3N-1$ and insulating otherwise\cite{dice-ribbon0}. It has been observed that in Haldane phases (with $\phi_h=\pi/2$), the spectra for 
$N_r=3N-1$ and $N_r\neq 3N-1$ are slightly different around the flat band, 
but the number and nature of edge states remain unaltered. The
edge states for nanoribbons with zigzag boundaries have also been analyzed and similar results as armchair were obtained.

\section{Haldane model in quantizing magnetic field}
Let the Haldane model be subjected to a uniform magnetic field ${\bf B}=B \hat{z}$. 
The vector potential ${\bf A}$ can be chosen in Landau gauge $(0,B x, 0)$ with $B>0$. 
We take the continuum model of massive Dirac Hamiltonian (\ref{ham-low}) near the Dirac points 
and incorporate the effect of magnetic field by minimal coupling 
${\bf p}\rightarrow({\bf p}+e{\bf A})$. Then, the Hamiltonian can be written as
\begin{equation}\label{ham-low-mag}
\begin{aligned}
\hat H_\mu=v_f\left(\mu \hat p_x S_x+(\hat p_y+eB \hat x) S_y\right)+m_\mu v_f^2 S_z.
\end{aligned}
\end{equation}
Since $[\hat p_y,\hat H_\mu]=0$, an eigenstate can be chosen as 
$|\lambda_\mu(x)\rangle=\e^{i q_y y}|\psi_\mu(x)\rangle$, 
where $|\psi_\mu(x)\rangle=(a\phi_a^\mu(x)\hspace{.3cm}b \phi_b^\mu (x) 
\hspace{.3cm}c\phi_c^\mu(x))^T$. Substituting $|\lambda_\mu(x)\rangle$ in 
the Schrodinger equation, we get $H_\mu |\psi_\mu(x)\rangle=\epsilon_\mu |\psi_\mu(x)\rangle$ with
\begin{equation}
H_\mu=\left(\begin{array}{ccc}
\epsilon_m & \epsilon_B f_\mu (\hat{x},\hat{p_x}) & 0\\
\epsilon_B f_\mu^\dagger (\hat{x},\hat{p_x}) & 0 & \epsilon_B f_\mu (\hat{x},\hat{p_x})\\
0 & \epsilon_B f_\mu^\dagger (\hat{x},\hat{p_x})& -\epsilon_m
\end{array}\right),
\end{equation}
where $\epsilon_m^\mu = m_\mu v_f^2$, $\epsilon_B = v_f\sqrt{\hbar e B}, 
f_\mu (\hat{x},\hat{p_x}) = -i(\delta_{\mu,1} \hat{a} + \delta_{\mu,-1} \hat{a}^\dagger)$. 
Here, $\hat{a}$ and $\hat{a}^\dagger$ are the lowering and raising operators of simple harmonic 
oscillator defined as
\begin{equation}
\hat{a} = \sqrt{\frac{eB}{2\hbar}} 
\l[\l(\hat{x}+\frac{\hbar q_y}{e B}\r) +i\frac{\hat{p}_x}{e B}\r]
\end{equation} 
and $\hat{a}^\dagger$ can be obtained by taking complex conjugate
of $\hat{a}$.
It turns out that the eigenspinor should be of the form
\begin{equation}\label{eigenstates}
|\psi_n^\mu(x)\rangle=
\left(\begin{array}{c}
a_n^\mu\l(\delta_{\mu,1} u_{n-1}(x-X)+\delta_{\mu,-1} u_{n+1}(x-X)\r)\\
b_n^\mu u_{n} (x-X)\\
c_n^\mu\l(\delta_{\mu,1} u_{n+1}(x-X)+\delta_{\mu,-1} u_{n-1}(x-X)\r)
\end{array}\right)
\end{equation}
for $n\geq1$, where $u_n(z)$ is eigenfunction of the $n$th level of harmonic oscillator 
and $X=-\hbar q_y/(eB)$. Using $\hat{a} u_{n} (x-X)= \sqrt{n} u_{n-1} (x-X) $ 
and $\hat{a}^\dagger u_{n} (x-X)= \sqrt{n+1} u_{n+1} (x-X) $ , we get the 
characteristic equation in eigenvalues $\epsilon$ as
\begin{equation}\label{depressed}
\epsilon^3-\epsilon\left(\epsilon_m^2+(2n+1)\epsilon_B^2\right)+\mu\epsilon_m^\mu \epsilon_B^2 =0.
\end{equation}

\begin{figure}[htbp]
\hspace{-0.4cm}\includegraphics[trim={0cm 0cm 0cm 0cm},clip,width=9cm]{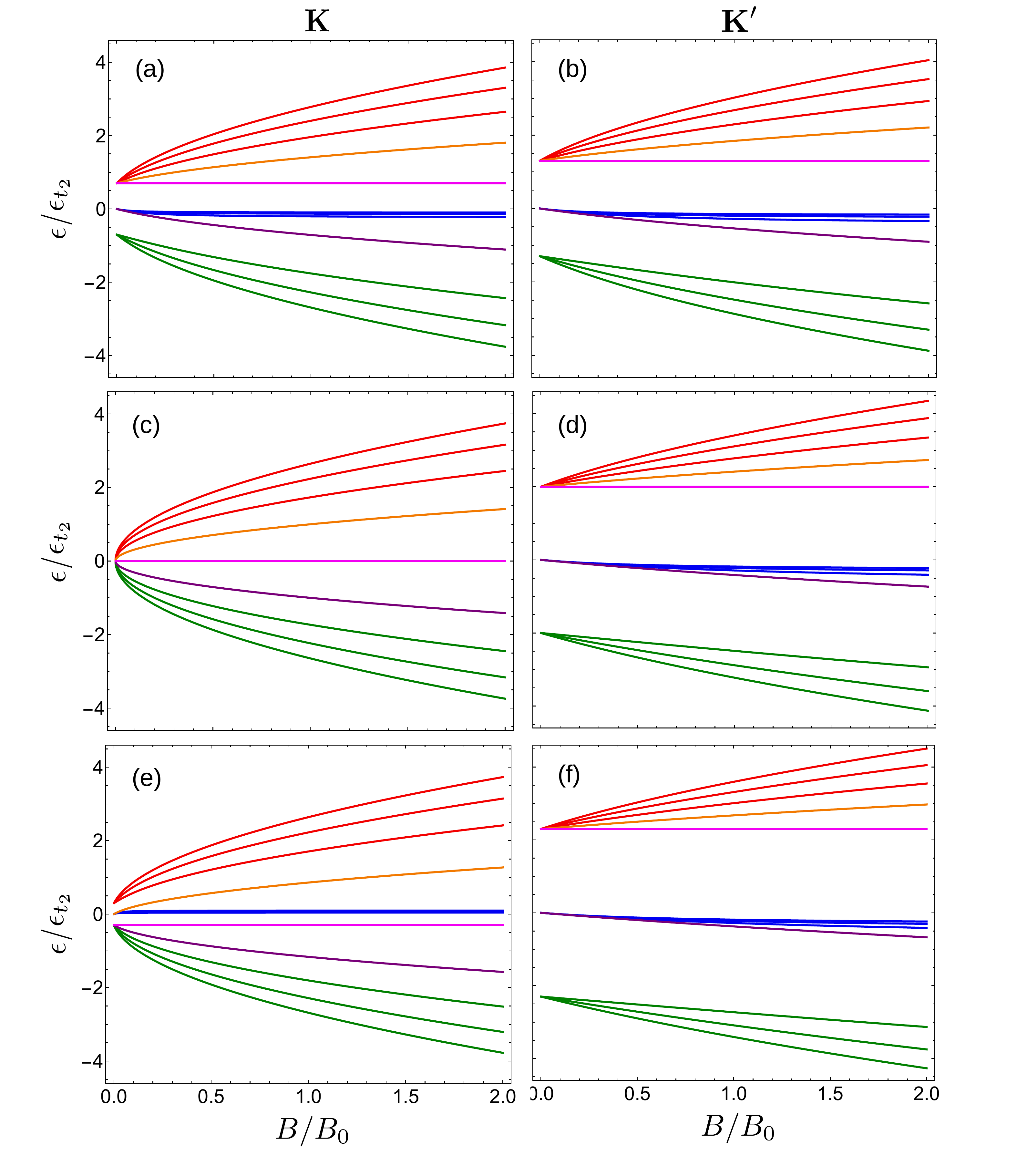}
\caption{Variation of Landau level energies around ${\bf K}$ and ${\bf K^\prime}$ points 
with magnetic field for $\gamma=0.3$ [(a),(b)], $\gamma=1.0$ [(c),(d)] and $\gamma=1.3$ [(e),(f)].} 
\label{landaulevelnew}
\end{figure}

Equation (\ref{depressed}) has the form of a depressed cubic equation
\begin{equation}
\epsilon^3 + \alpha \epsilon + \beta = 0
\end{equation}. 
Its solutions are given by
\begin{equation}\label{cubic-roots}
\epsilon_{nj}^\mu = 2\sqrt{\frac{-\alpha}{3}} \cos\Big[ \frac{1}{3} 
\cos^{-1}\bigg(\frac{3\beta}{2\alpha}\sqrt{\frac{-3}{\alpha}}\bigg)-\frac{2\pi j}{3} \Big]
\end{equation}
with $\alpha=-\l[(\epsilon_m^\mu)^2+(2n+1)\epsilon_B^2\r]$, $\beta=\mu\epsilon_m^\mu \epsilon_B^2$ 
and $j=0,1$ and $2$. For $n\geq1$, the eigenvalues or Landau level energies of each valley 
$\epsilon_{nj}^\mu$ are given by equation (\ref{cubic-roots}). 

The components of the eigen spinors $\psi_{nj}^\mu$ are given by
\begin{equation}
a_{nj}^\mu=-i\l(\frac{\delta_{\mu,1}\sqrt{n B/B_0} + 
\delta_{\mu,-1}\sqrt{(n+1) B/B_0}}{\tilde{\epsilon}_{nj}^\mu-(\gamma-\mu)}\r)b_{nj}^\mu,
\end{equation}
\begin{equation}
c_{nj}^\mu=i\l(\frac{\delta_{\mu,1}\sqrt{(n+1) B/B_0} + 
\delta_{\mu,-1}\sqrt{n B/B_0}}{\tilde{\epsilon}_{nj}^\mu+(\gamma-\mu)}\r)b_{nj}^\mu,
\end{equation}
and
\begin{equation}
\begin{aligned}
|b_{nj}^\mu|&^2  =\delta_{\mu,1}\l( 1+ \frac{n B/B_0}{(\tilde{\epsilon}_{nj}^\mu-(\gamma-\mu))^2} 
+ \frac{(n+1) B/B_0}{(\tilde{\epsilon}_{nj}^\mu + (\gamma-\mu))^2}\r) \\ 
& +\delta_{\mu,-1}\l( 1+ \frac{(n+1) B/B_0}{(\tilde{\epsilon}_{nj}^\mu-(\gamma-\mu))^2} + 
\frac{n B/B_0}{(\tilde{\epsilon}_{nj}^\mu+(\gamma-\mu))^2}\r),
\end{aligned}
\end{equation}
where $\tilde{\epsilon}_{nj}^\mu=\epsilon_{nj}^\mu/\epsilon_{t_2}$ and 
$B_0=\epsilon_{t_2}^2/(e\hbar v_f^2)$ is a magnetic field scale of the system.

Two other possible eigenspinors are 
$|\psi_0^\mu(x)\rangle=\left(\delta_{\mu,-1}a_0^\mu u_{1}(x-X) 
\hspace{.3cm}b_0^\mu u_{0}(x-X)\hspace{.3cm}\delta_{\mu,1}a_0^\mu u_{1}(x-X)\right)^T$ 
and $|\psi_{00}(x)\rangle=\l( \delta_{\mu,-1}u_{0}^\mu(x-X) 
\hspace{.3cm}0\hspace{.3cm}\delta_{\mu,1}u_{0}^\mu(x-X)\r)^T$ 
with energies $\epsilon_0^\mu=(-\mu \epsilon_m^\mu\pm\sqrt{(\epsilon_m^\mu)^2+ 
4\epsilon_B^2})/2$ and $\epsilon_{00}^\mu=-\mu \epsilon_m^\mu$ respectively. 
The amplitudes $a_0^\mu$ and $b_0^\mu$ are given by
\begin{equation}
a_0^\mu=\l(\frac{i\l(\delta_{\mu,1}-\delta_{\mu,-1}\r)\tilde{\epsilon}_0^\mu}{\sqrt{B/B_0}}\r)b_0^\mu, 
\hspace{0.5cm} |b_0^\mu|=\l[1+ \frac{(\tilde{\epsilon}_0^\mu)^2}{B/B_0}\r]^{-1/2},
\end{equation}
where $\tilde{\epsilon}_0^\mu=\epsilon_0^\mu/\epsilon_{t_2}$.

The variation of Landau level energies with magnetic field is shown in Fig. \ref{landaulevelnew} 
for the two valleys. The valley-symmetry of the spectrum is broken for $\gamma\neq0$. 
For no mass term (Semenoff or Haldane-like) in the Hamiltonian, we get infinite number of 
degenerate zero energy Landau levels \cite{Illes1,Tutul}. The Haldane mass term splits 
all these levels shifting them towards positive or negative energy in each valley, as 
shown by blue curves in the figure. This was observed for massive dice lattice as well \cite{duan}.
In each valley, there exists a constant energy level 
$\epsilon_{00}^\mu$ denoted by pink lines, whose magnitude is equal to the magnitude of 
the mass term in the respective valleys. For $\gamma\neq1$, the Landau levels in 
$K$ valley vary nearly as $\sim\sqrt{B}$ for $\epsilon_m\ll \epsilon_B$ 
[Figs. \ref{landaulevelnew}(a),(e)]. For $\gamma=1$, the energies scale exactly as 
$\sqrt{B}$ at $K$ valley [Fig. \ref{landaulevelnew}(c)], where the gap closes with massless 
spin-1 Dirac-Weyl dispersion. The mass term in the $K^\prime$ valley is large for 
$\gamma\geq1$. In this case, the spectrum varies nearly as $\sim B$ for 
$\epsilon_B\ll\epsilon_m$ [Fig. \ref{landaulevelnew}(d),(f)].

\section{Longitudinal conductivity}
Using the Kubo formalism, the longitudinal conductivity $\sigma_{xx}$ is obtained as 
(see Appendix \ref{app-sigmaxx} for derivation)
\begin{equation}\label{sigmaxx}
\sigma_{xx} = \tilde \sigma_0 \sum_{n,j,\mu}^{}I_{nj}^\mu 
f_{nj}^\mu(\epsilon_{n, j}^\mu) \l \{ 1 - f_{nj}^\mu(\epsilon_{n, j}^\mu) \r \},
\end{equation}
where
$ \tilde \sigma_0 = (g_s e^2 n_{\rm im} V_0^2)/(\pi h\Gamma_0 k_B T l_0^2)$, $f_{nj}^\mu 
= (\e^{(\epsilon_{nj}^\mu-E_f)/K_BT}+1)^{-1}$.  
Also, the term $I_{nj}^\mu$ is obtained as
\begin{equation} \label{Inj}
\begin{aligned}
I_{nj}^\mu  = & |a_{nj}^\mu|^4 [\delta_{\mu,1}(2n-1) + \delta_{\mu,-1}(2n+3) ] 
+ |b_{nj}^\mu|^4 (2n+1) \\ 
& + |c_{nj}^\mu|^4 [\delta_{\mu,1}(2n+3) + \delta_{\mu,-1}(2n-1)] 
- 2 |a_{nj}^\mu|^2|b_{nj}^\mu|^2 \\ 
& \times [\delta_{\mu,1}n + \delta_{\mu,-1}(n+1) ] - 2|b_{nj}^\mu|^2|c_{nj}^\mu|^2 
[\delta_{\mu,1}(n+1)  \\ & + \delta_{\mu,-1}n].
\end{aligned}
\end{equation}

\begin{figure}[htbp]
\hspace{-0.4cm}\includegraphics[trim={0cm 0cm 0cm 0cm},clip,width=9cm]{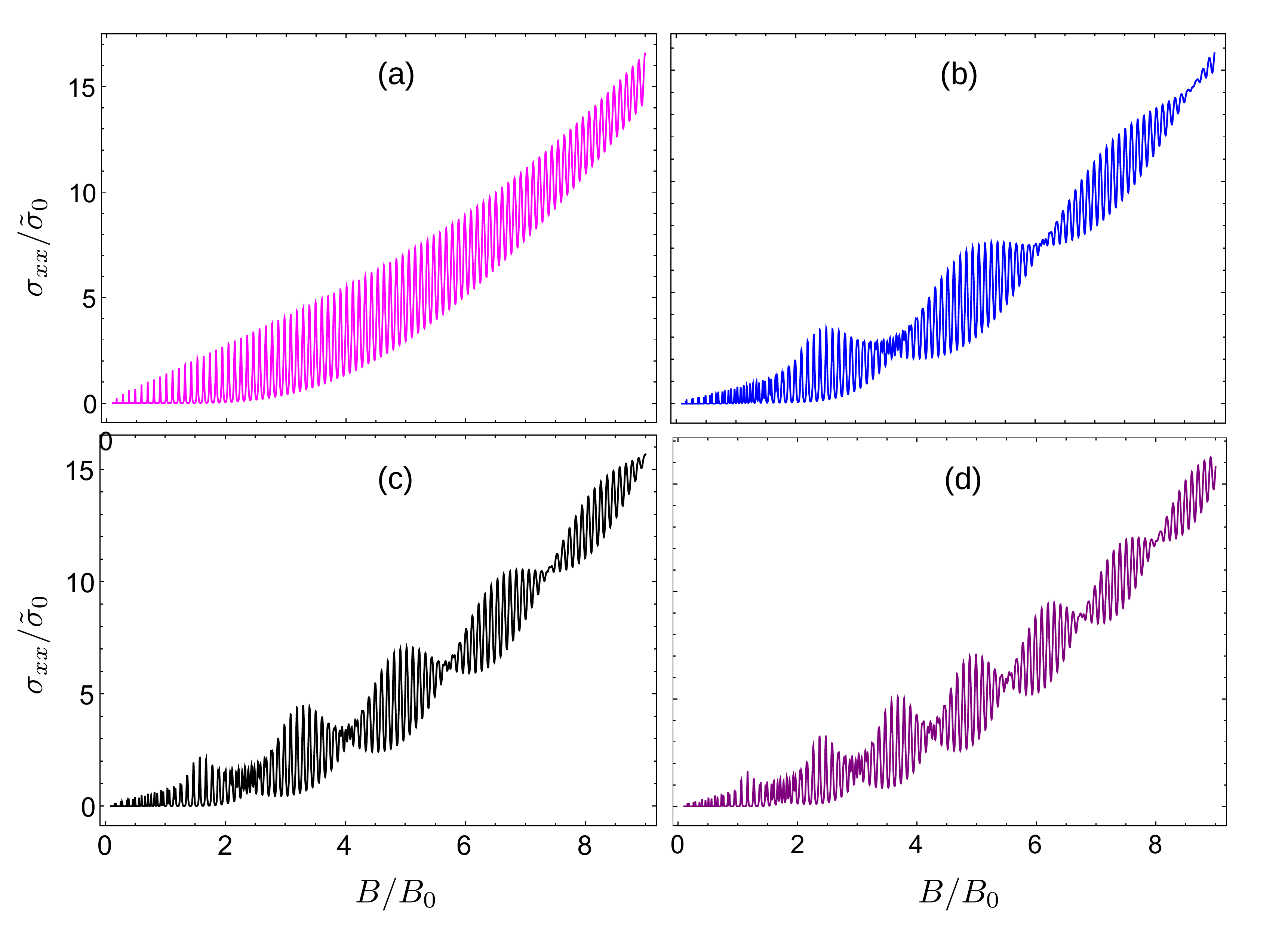}
\caption{Longitudinal conductivity ($\sigma_{xx}$) of Haldane model of dice lattice as a function 
of inverse magnetic field ($1/B$) for (a) $\gamma=0$, (b) $\gamma=0.2$, (c) $\gamma=0.3$ 
and (d) $\gamma=0.4$. Other parameters are $E_f=4\epsilon_{t_2}$ and 
$k_BT=0.005 \epsilon_{t_2}$. }
\label{beats}
\end{figure}

\begin{figure}[htbp]
\hspace{-0.4cm}\includegraphics[trim={0cm 0cm 0cm 0cm},clip,width=7cm]{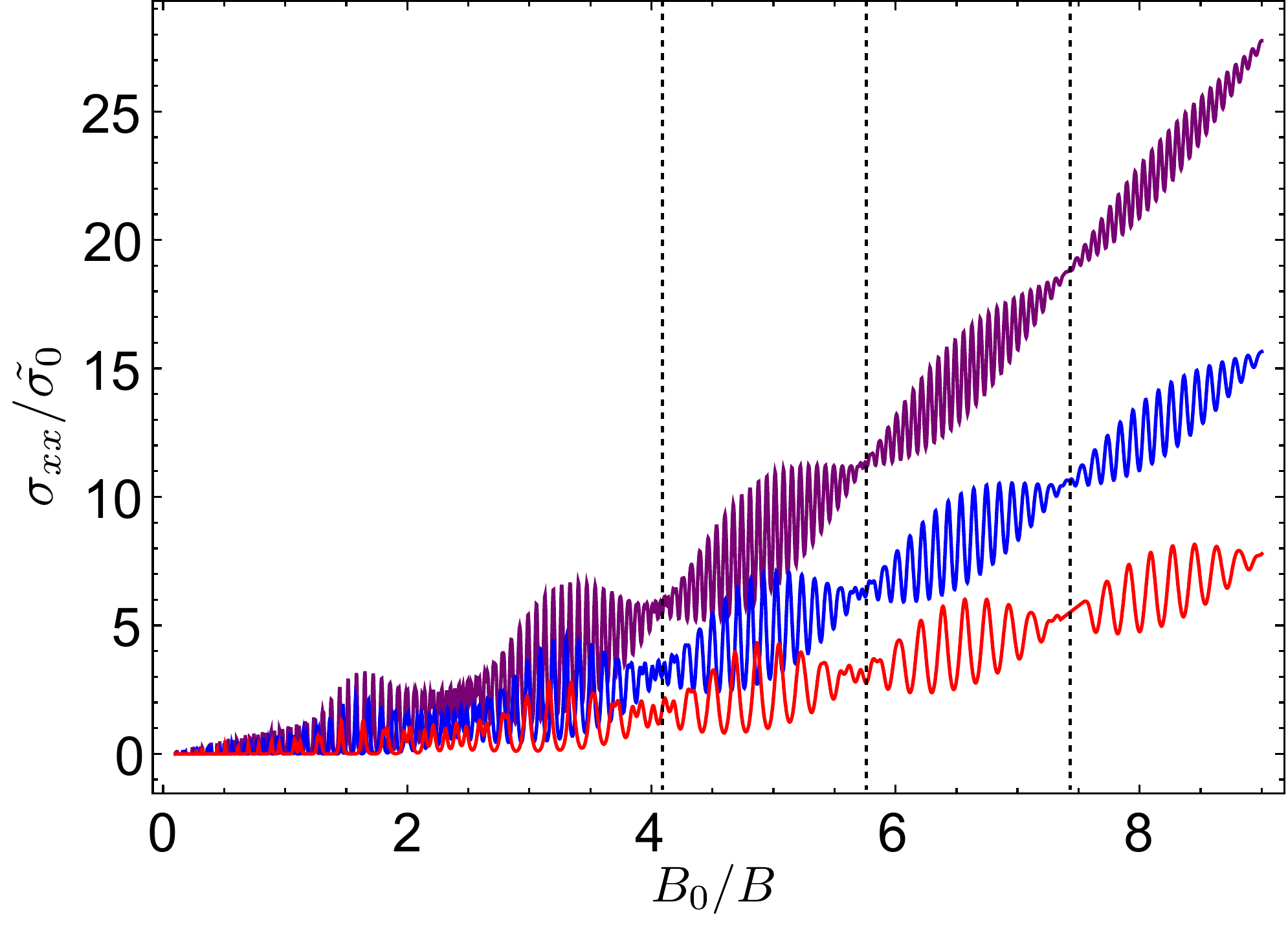}
\caption{Longitudinal conductivity ($\sigma_{xx}$) of Haldane model of dice lattice as a 
function of inverse magnetic field ($1/B$) for $\epsilon_f = 3.5$ (red), 
$\epsilon_f = 4.5$ (blue) and $\epsilon_f = 5.5$ (purple). Other parameters are 
$\gamma=0.3$ and $k_BT=0.005 \epsilon_{t_2}$.}
\label{beats2}
\end{figure}
Substituting Eq. (\ref{Inj}) into Eq. (\ref{sigmaxx}), we obtain the longitudinal 
conductivity as a function of $B$. 
In Fig. \ref{beats}, $\sigma_{xx}/\tilde \sigma_0$ is plotted as a function of $B_0/B$ for 
different values of semenoff mass $M$ at a high $E_f$ in the conduction band. We get the 
usual SdH oscillations in $\sigma_{xx}$ for $M=0$. For finite $M$, beats appear in the SdH 
oscillations and the frequency of beats increases with $M$. In Fig. \ref{beats2}, the beating 
pattern is plotted for different values of $E_f$ for a given value of $M$. It is observed the 
number of oscillations between two nodes increases with $E_f$, but the beat frequency is 
apparently constant. Similar phenomena occurs when $E_f$ lies in the valence band.

To qualitatively explain the nature of the plots in Fig. \ref{beats},\ref{beats2}, 
we consider an approximated formula of SdH oscillations in 2D electron system at low 
temperatures and low magnetic fields, which is given by \cite{sdh}
\begin{equation}
\frac{\sigma_{xx}}{\sigma_0} = 1 - \lambda \sum_{\mu}^{} 
\frac{2\pi^2 k_B T/\epsilon_B}{\sinh(2\pi^2 k_B T/\epsilon_B)} 
\e^{-\frac{\pi \hbar}{\tau \epsilon_B}}\cos\l[\frac{\hbar S_F^\mu}{eB}+\phi_0^\mu\r],
\end{equation}
where $\lambda$ is a constant, $\epsilon_B = \hbar\omega_c$, 
$\phi_0^\mu$ is the energy- and valley-dependent Berry phase and 
$ S^\mu_F = \pi (k_f^\mu)^2$ is the area enclosed by the Fermi circle 
in a given valley $\mu$. 
For massive dice lattice, $\phi_0^\mu = \mu(2\pi m_\mu v_f^2/E_f)$.  
Using Eq. (\ref{low}), we have $S_F^\mu = \pi \l(E_f^2-(m_\mu v_f^2)^2\r)/(\hbar v_f)^2$. 
The cosine terms of the two valleys act as harmonics with $S_F^\mu$ being the corresponding 
frequencies. The beats arise due to small difference in $S_F^+$ and $S_F^-$ due to 
difference in magnitude of mass terms in the two valleys. We can obtain the beat frequency 
by modelling the longitudinal concutivity as
\begin{equation}
\frac{\sigma_{xx}}{\sigma_0}\sim\cos\l( \frac{\hbar S_F^+}{eB}+\phi_0^+\r) + 
\cos\l( \frac{\hbar S_F^-}{eB}+\phi_0^-\r).
\end{equation}
On simplification, we obtain
\begin{equation}\label{coscos}
\frac{\sigma_{xx}}{\sigma_0}\sim\cos\l(2\pi\l[f_m\l(\frac{B_0}{B}\r)-\frac{1}{\epsilon_f}\r]\r) 
\cos\l(2\pi \gamma\l[\frac{B_0}{B}+\frac{1}{\epsilon_f}\r]\r),
\end{equation}
where $\epsilon_f = E_f/\epsilon_{t_2}$ and $f_m=(\epsilon_f^2-\gamma^2-1)/2$ is the frequency of 
modulation. The second cosine factor in Eq. (\ref{coscos}) gives the beating envelope with 
beat frequency $f_b=2\gamma$. The position of $j$-th beating node is 
$(B_0/B)_j=(2j-1)/4\gamma-1/\epsilon_F$ where $j=1,2,3...$ . The interval between two 
successive beating nodes is $\Delta = (B_0/B)_{j+1}-(B_0/B)_j=1/f_b$.  
The number of oscillations between two successive nodes is given by 
$N = f_m\Delta=(\epsilon_F^2-\gamma^2-1)/(4\gamma)$.

From Fig. \ref{beats2}, we have $(B_0/B)_4 = 5.72$ and $(B_0/B)_5 = 7.39$ which gives 
$\Delta_{plot} = 1.67$. This exactly matches with the time period of beats given by 
$\Delta= 1/(2\gamma)$. Also, in the SdH plot for $\epsilon_f = 4.5$ in Fig. \ref{beats2},
we get nearly 16 oscillations between two successive nodes. This matches with the result 
$(=15.96)$ obtained from the expression for $N$.

The average frequency of oscillations arising from the two valleys is proportional to 
$E_f^2$. Thus, $N$ falls rapidly as $E_f$ approaches lower Landau levels of conduction or 
valence band and the beats gradually become indistinct. This is expected because the 
formation of beats requires the individual frequencies of the superposing harmonics to be 
much larger than their difference. So, well defined beats can be observed only when $E_f$ is 
large enough. In our analysis, we have chosen $E_f=4.5 \epsilon_{t_2}$ which is close to 
$\sim 50^{th}$ Landau level of the conduction band at either valley for $\gamma=0.3$ and  
$B_0/B=5$. Similar beats are also expected for Haldane model of graphene.

It is to be noted that although the Landau levels are not valley-degenerate 
in massive dice lattice as well \cite{duan}, it does not show beats in SdH oscillations.

\section{Conclusion}
We have constructed a theoretical Haldane-like model of dice lattice and investigated its 
topological properties within the tight-binding formalism. The phases of the system are 
dictated by the Semenoff mass, second neighbour hopping and periodic magnetic flux. Unlike 
the Haldane model of graphene which hosts phases representing a semi-metal, trivial insulator 
and a topogical insulator, this sytem supports a metalic phase in addition to the former. 
The metalic phase arises due to distortion of the flat band and its indirect overlap with 
either of the two other bands. These phases also gapped bands which may be topologically 
trivial or non-trivial. Chiral edge states show up in the band structure of a nanoribbon 
of the sytem in the non-trivial regime. A haldane phase with pure imaginary hoppings restores 
the dispersionless flat band. The Chern numbers of the bands are $-2,0$ and $2$ in the topological 
phases implying that the system may exhibit QAHE with two chiral edge channels. 
Exact expressions of Landau levels are derived from low-energy massive pseudospin-1 
Dirac Hamiltonians around the two Dirac points. Peculiar beating pattern appears in the 
SdH oscillations of magneto-conductivity when the magnitude of mass terms in the two Dirac 
valleys are unequal and filling is close to the higher Landau levels of the conduction or 
valence band. The information about the phase-determining parameters of the system such as 
Semenoff mass, next neighbour hopping and Fermi energy can be extracted from the beat 
frequency and the number of oscillations between two successive beating nodes.

\begin{center}
{\bf ACKNOWLEDGEMENTS}
\end{center}
We would like to thank Sonu Verma and Ritajit Kundu for
useful discussions.

\appendix{}

\section{Energy bands of Haldane model} \label{app-bands}
In this appendix, we present derivation of the energy bands of the Haldane
model for dice lattice. 
The Hamitonian (\ref{hamfinal}) yields the folowing characteristic equation of 
eigenvalues $\epsilon$ --
\begin{equation}\label{charac}
\epsilon^3-2p_0 \epsilon^2 - (p_z^2 - p_0^2 + 2|p_{xy}|^2)\epsilon + 2 p_0 |p_{xy}|^2 = 0,
\end{equation}
where $p_0 =2 t_2 h_0({\bf k}) \cos\phi_h$, $p_z = M - 2 t_2 h_z({\bf k}) 
\sin\phi_h$ and $p_{xy} = t\l(g_x({\bf k})-ig_y({\bf k})\r)/\sqrt{2}$. 
Solutions of this equation gives the band structure of the sytsem as functions of 
$M, t_2$ and $\phi_h$.

Equation (\ref{charac}) has the form
\begin{equation}\label{charac-abcd}
A\epsilon^3 + B\epsilon^2 + C\epsilon + D = 0,
\end{equation}
with $A=1, B=-2p_0, C=-(p_z^2-p_0^2+2|p_{xy}|^2)$ and $D=2 p_0 |p_{xy}|^2$. 
The solutions can be obtained by converting it to a {\it depressed} cubic equation. 
Substituting $\epsilon = \omega - B/3A$ in Eq. (\ref{charac-abcd}) and dividing 
by $A$, we get
\begin{equation}\label{charac-pq}
\omega^3 + p \omega + q = 0,
\end{equation}
where
\begin{equation}
p = \frac{3AC-B^2}{3A^2}, \hspace{0.2cm}
q = \frac{2B^3-9ABC+27A^2D}{27A^3}.
\end{equation}
Equation (\ref{charac-pq}) has the form of a {\it depressed} cubic equation with $p<0$ 
for all values of ${\bf k}$ in our system. Since all the eigenvalues are real, 
the solutions are of trigonometric form --
\begin{equation}\label{roots-appendix}
\omega_j = 2\sqrt{\frac{-p}{3}} \cos\Big[ \frac{1}{3} 
\cos^{-1}\bigg(\frac{3q}{2p}\sqrt{\frac{-3}{p}}\bigg)-\frac{2\pi j}{3} \Big], \hspace{0.2cm}j=0,1,2.
\end{equation}
The energy bands of the Haldane model of the dice lattice are given by 
$\epsilon_j = \omega_j - B/3A$.

\section{Magnetoconductivity from the Kubo formula} \label{app-sigmaxx}
Here we will provide the derivation of the analytical expression of the longitudinal 
conductivity in the linear response regime where the electric field is very weak.
For this purpose we will be using the well-known Kubo formalism \cite{Kubo}. 
In general, the longitudinal conductivity has diffusive and collisional contributions. 
In presence of quantizing magnetic field, the diffusive contribution exactly vanishes 
since the diagonal elements of the velocity operator are simply zero. 
Therefore, the longitudinal conductivity solely arises due to the 
collisional process. 

In the framework of Kubo formalism, the general expression for the collisional
conductivity in presence of quantizing magnetic field is given 
by \cite{vasilo,Kubo_coll1,Kubo_coll3,Kubo_coll4}
\begin{eqnarray}
\sigma_{xx} = \frac{e^2}{S k_BT}
\sum_{\xi, \xi^\prime} f(\epsilon_\xi)\{1-f(\epsilon_{\xi^\prime})\}
W_{\xi \xi^\prime} (x_\xi - x_{\xi^\prime})^2,
\end{eqnarray}
where $S$ is the surface area of the system, 
$\xi \equiv(j, n, q_y, \mu)$ represents a set of all quantum numbers, 
$T$ being the temperature of the system, 
$x_\xi = \langle \xi \vert x \vert \xi\rangle = q_yl_0^2$, and
$f(\epsilon_\xi) = [\e^{(\epsilon_\xi - E_f)/k_BT} + 1]^{-1}$ is the Fermi-Dirac 
distribution function. 
Moreover, $W_{\xi \xi^\prime}$ describes the probability that an electron 
makes a transition from an initial state $\vert \xi\rangle$ to a final state 
$\vert \xi^\prime\rangle$. Its expression for elastic scattering by static impurities 
is given by
\begin{eqnarray}\label{Trans_Prob}
W_{\xi,\xi^\prime}=\frac{2\pi n_{\rm im}}{\hbar S}
\sum_{\bf k} \vert V({\bf k})\vert^2 \vert F_{\xi, \xi^\prime}\vert^2
\delta(\epsilon_\xi-\epsilon_{\xi^\prime}),
\end{eqnarray}
where $n_{\rm im}$ is the impurity density and 
$V({\bf k})$ is the Fourier transform of the screened Coulomb 
potential $V({\bf r}) = e^2 \e^{-k_sr}/(4\pi\epsilon_0\epsilon_r r)$ with
$\epsilon_0$ is the free space permittivity, 
$\epsilon$ is the dielectric constant of the medium and $k_s$ is the screened 
wave vector.
The expression of $V({\bf k})$ for a 2D system is 
$V({\bf k})=e^2/(4\pi\epsilon_0\epsilon_r\sqrt{k^2 + k_s^2})$.
Finally, $F_{\xi, \xi^\prime}$ denotes the form factor which is defined as 
$F_{\xi, \xi^\prime}=\langle \xi^\prime\vert e^{i{\bf k} \cdot {\bf r}}\vert\xi\rangle$. 
We now consider only the intra-band ($ j^\prime = j$) and intra-level 
($n^\prime=n$) scattering because of the presence of the term 
$\delta(\epsilon_\xi-\epsilon_{\xi^\prime})$ in Eq. (\ref{Trans_Prob}). 
The valley-dependent form factor is simplified as
\begin{equation}
\begin{aligned}\label{Fn}
|F_{nj}^\mu (u)| = & [|a_{nj}^\mu|^2 (\delta_{\mu,1}L_{n-1}(u) + 
\delta_{\mu,-1}L_{n+1}(u))  \\ 
& + |b_{nj}^\mu|^2 L_n(u) + |c_{nj}^\mu|^2 (\delta_{\mu,1}L_{n+1}(u) \\ 
& + \delta_{\mu,-1}L_{n-1}(u))]\e^{-u/2},
\end{aligned}
\end{equation}
where $L_n(u)$ is the Laguerre polynomial of order $n$.


The sharp Landau levels are broadened by the static impurities present in the system: 
$\delta(\epsilon_\xi-\epsilon_{\xi^\prime})=
(1/\pi)\Gamma_0/[(\epsilon_\xi-\epsilon_{\xi^\prime})^2+\Gamma_0^2]$
with $\Gamma_0$ being the broadening parameter. 
It may be written as 
$\delta(\epsilon_\xi-\epsilon_{\xi^\prime})\simeq 1/(\pi\Gamma_0)$
for intra-level and intra-band scattering.
Further, $V({\bf k})$ is approximated as 
$V({\bf k})\simeq e^2/(4\pi\epsilon_0\epsilon k_s)\equiv V_0$ since
small values of $k^2$ will be contributing more due to the presence
of exponentially decaying term $e^{-u}$ in the expressions of
$\big\vert F_{n, j}^\mu \big\vert^2$. 

Using the fact that $\sum_{q_y} \rightarrow g_s S/(2\pi l_0^2)$ with $g_s$ 
being the spin-degeneracy and 
$\sum_{\bf k}\rightarrow S/(2\pi)^2\int k\, dk\, d\theta$ with $\theta$ being the polar 
angle of ${\bf k}$, we finally obtain the following expression for the longitudinal
conductivity as 
\begin{eqnarray}\label{coll_cond1}
\sigma_{xx} & = & \tilde \sigma_0
\sum_{j, \mu, n} I_{n,j}^{\mu}  f\big(\epsilon_{n, j}^\mu \big)
\l \{1 - f\big(\epsilon_{n, j}^\mu \big) \r\},
\end{eqnarray}
where $ \tilde \sigma_0 = (g_s e^2  n_{\rm im} V_0^2)/(\pi h \Gamma_0 k_BTl_0^2)$
and $I_{n,j}^{\mu} = 
\int_0^\infty u \vert F_{n j}^\mu (u) \vert^2 \, du$.
Using the standard results of $\int_{0}^{\infty}L_n^2(u)\e^{-u}udu=2n+1$ and 
$\int_{0}^{\infty}L_n(u)L_{n-1}(u)\e^{-u}udu=-n$,
$I_{n,j}^{\mu} $ is obtained as  
\begin{equation}
\begin{aligned}
I_{n,j}^{\mu} & =  |a_{nj}^\mu|^4 [\delta_{\mu,1}(2n-1) + \delta_{\mu,-1}(2n+3) ]
+ |b_{nj}^\mu|^4 (2n+1) \nn \\
& +  |c_{nj}^\mu|^4 [\delta_{\mu,1}(2n+3) + \delta_{\mu,-1}(2n-1)]
- 2 |a_{nj}^\mu|^2|b_{nj}^\mu|^2 \nn \\
& \times [\delta_{\mu,1}n + \delta_{\mu,-1}(n+1) ] - 2|b_{nj}^\mu|^2|c_{nj}^\mu|^2
[\delta_{\mu,1}(n+1) \nn \\ 
& + \delta_{\mu,-1}n].
\end{aligned}
\end{equation}


\end{document}